\documentclass[aps,prb,twocolumn]{revtex4-1}
\usepackage{amsmath}
\usepackage{amssymb}
\usepackage{natbib}
\usepackage{physics}
\usepackage{hyperref}
\usepackage{graphicx}
\usepackage{color}
\usepackage{ulem}

\begin{document}
	\title{Meissner Currents Induced by Topological Magnetic Textures in Hybrid Superconductor/Ferromagnet Structures}
	\author{Samme M. Dahir, Anatoly F. Volkov, and Ilya M. Eremin}
	\affiliation{Institut f\"ur Theoretische Physik III,
		Ruhr-Universit\"{a}t Bochum, D-44780 Bochum, Germany}
	\date{\today }

	\begin{abstract}
		Topological spin configurations in proximity to a superconductor have recently attracted great interest due to the potential application of the former in spintronics and also as another platform for realizing non-trivial topological superconductors. Their application in these areas requires precise knowledge of the existing exchange fields and/or the stray-fields which are therefore essential for the study of these systems.
		Here, we determine the effective stray-field \(\vb{H}_{str}\) and the Meissner currents \(\vb{j}_{\text{S}}\) in a Superconductor/Ferromagnet/Superconductor (S/F/S) junction produced by various nonhomogenous magnetic textures \(\vb{M}(\vb*{r})\) in the F. The inhomogeneity arises either due to a periodic structure with flat domain walls (DW) or is caused by an isolated chiral \mbox{magnetic skyrmion (Sk)}. 
		We consider both Bloch-- and \mbox{N\'{e}el--type} Sk and also analyze in detail the periodic structures of different types of DW's--- that is \mbox{Bloch--type} DW (BDW) and \mbox{N\'{e}el--type} DW  (NDW) of finite width with in- and out-of-plane magnetization vector \(\vb{M}(x)\). The spatial dependence of the fields \(\vb{H}_{str}(\vb*{r})\) and Meissner currents \(\vb{j}_{\text{S}}(\vb*{r})\)  are shown to be qualitatively different for the case of Bloch-- and \mbox{N\'{e}el--type} magnetic textures. While the spatial distributions in the upper and lower S are identical for \mbox{Bloch--type} Sk and DW's they are asymmetric for the case of \mbox{N\'{e}el--type} magnetic textures.
		The depairing factor, which determines the critical temperature \(T_{c}\) and which is related to vector potential of the stray-field, can have its maximum at the center of a magnetic domain but also, as we show, above the DW. For Sk's the maximum is located at a finite distance within the Sk radius \(r_{\text{Sk}}\). Based on this, we study the nucleation of superconductivity in the presence of DW's. Because of the asymmetry for \mbox{N\'{e}el--type} structures, the critical temperature \(T_{c}\) in the upper and lower S is expected to be different. The obtained results can also be applied to S/F bilayers.
 	\end{abstract}
	
	\maketitle
	
	\date{\today}
Over the past decades, continuous efforts have been made to study superconductor-ferromagnet heterostructures due to a variety  of interesting features caused by the proximity effect, \textit{i.e.}, the penetration of Cooper pairs from the superconductor (S) into the ferromagnet (F). The most interesting and well established  effects are the sign reversal of the Josephson current in S/F/S junctions and the appearance of a long-ranged triplet component (see review articles \cite{golubov_current-phase_2004,buzdin_proximity_2005,bergeret_odd_2005,eschrig_spin-polarized_2015,linder_superconducting_2015,balatsky_odd_2017,ohnishi2020spintransport} and references therein).
	
Other interesting features involve the interplay of various types of topological defects that, under certain conditions, can be present in the superconductor and/or ferromagnet. One of these topological defects are the Abrikosov vortices which occur in \mbox{type--II} superconductors\cite{Abrikosov:1956sx} in the magnetic field interval, \mbox{\(H_{c1}<H_{\text{ext}}<H_{c2}\)}. 
There are also several different topological structures that can be found in ferromagnets. The most prominent ones are magnetic domain walls (DW), where the magnetization vector \(\vb{M}\) rotates by an angle \(\pi\) across the DW. Another example  of a topological defect that has received much attention recently due to its potential application in spintronics are the so-called magnetic Skyrmions (Sk)\cite{bogdanov1989thermodynamically,Leonov2016,fert_magnetic_2017,Everschor2018}. These local whirl-like structures are topologically equivalent to two DW's as one can map the inner part of the Sk on the stripes between two domains via conformal transformation. Similar to flat DW's, where the magnetization vector \(\vb{M}\) changes its direction by rotating either in the \((x,z)\)-plane (\mbox{N\'{e}el--type}) or in the \((y,z)\)-plane (\mbox{Bloch--type}), the winding of chiral Sk can either have a Bloch-- or a N\'{e}el--like structure. Which type of chiral Sk is realized depends on the underlying chiral interaction. Note that there is already some work on the mutual interaction between topological defects occurring in ferromagnets and superconductors, see review\cite{lyuksyutov_ferromagnetsuperconductor_2005} and references therein. In the absence of the direct proximity effect (no direct contact between S and F), this interaction is realized through the magnetic stray-field \(\vb{H}_{\text{str}}\)  generated by the non-uniform magnetic textures in the F and the magnetic field associated with the superconducting vortices. The creation of Pearl and Abrikosov vortices in S/F structures with and without DW's has been analyzed in Refs.\cite{lyuksyutov_magnetization_1998,erdin_topological_2001,lyuksyutov_ferromagnetsuperconductor_2005,milosevic_interaction_2003,laiho_penetration_2003}. More recently, the spontaneous creation of vortices in S/F structures with Sk's with and without direct proximity effect was also studied theoretically\cite{Hals2016,baumard_generation_2019,dahir_interaction_2019,Menezes2019,Rex2019,Palermo2020}
	
As it is well known, there is no stray-field \(\vb{H}_{\text{str}}\) outside of uniformly magnetized infinite film \cite{landau_electrodynamics_nodate}. However, within the ferromagnet the magnetic induction \(\vb{B}_{\text{F}}\) or the magnetic field \(\vb{H}_{\text{F}}\) can still acquire finite values, {\it i.e.}, \mbox{\(\vb{B}_{\text{F}}=4\pi \vb{M}_{0}\)} and \mbox{\(\vb{H}_{\text{F}}=0\)} for the in-plane magnetization and \mbox{\(\vb{B}_{\text{F}}=0\)} and \mbox{\(\vb{H}_{\text{F}}=-4\pi \vb{M}_{0}\)} for the out-of-plane magnetization. Therefore for a uniform magnetization \(\vb{M}_{0}\) in the F of a S/F/S structure, both the \(\vb{B}_{\text{S}}\), the \(\vb{H}_{\text{S}}\) and the Meissner current \(\vb{j}_{\text{S}}\) are equal zero in the superconductors  where \mbox{\(\vb{B}^{\text{S}}=\vb{H}_{\text{S}}\)}. Thus, non-zero stray-fields and Meissner currents can only occur if the magnetization of an infinite F is non-homogeneous. This was studied in S/F structures with DW's of zero width in Refs.\cite{laiho_penetration_2003,stankiewicz_magnetic_1997,bulaevskii_ferromagnetic_2000}, for DW's of finite width in Ref.\cite{Burmistrov2005} and for a magnetic vortex configuration in Ref.\cite{Frearman_2005}. In the presence of a proximity effect and spin-orbit coupling, the Meissner current was recently calculated in a bilayer S/F structure with a particular \mbox{N\'{e}el--type} Sk in the F and a vortex in the S \cite{baumard_generation_2019}.
	
Despite of existing literature, there are still no systematic studies of Meissner currents in S/F and S/F/S structures with different topological magnetic textures (Sk's or flat DW's) with different orientations of the magnetization vector \(\vb{M}\). This is particular interesting due to potential realization of Majorana fermions in such heterostructures\cite{Yang2016,Garnier2019,Rex2019}. In the present paper we address this topic, by analyzing S/F/S systems with an isolated magnetic Sk (Bloch-- and \mbox{N\'{e}el--type} Sk) or with a periodic flat DW structure (out-of-plane and in-plane magnetization, Bloch and N\'{e}el DW's) in the ferromagnetic material.

Assuming that there is no proximity effect present and that there are no Abrikosov vortices in the S/F/S structure , \textit{i.e.}, magnetic stray-fields are supposed to be less than the critical field \(H_{c1}\), we find the effective magnetic stray-field from which we deduce the induced screening currents \(\vb{j}_{\text{S}}\). Note that the obtained spatial distribution of the current density \(\vb{j}_{\text{S}}\) in the S in a Josephson system S/F/S is qualitatively similar to that in S/F bilayer. 
The knowledge of the effective stray-field \(\vb{H}_{\text{str}}\) and the current density \(\vb{j}_{\text{S}}\) allows one to estimate the region where superconductivity nucleates upon decreasing the temperature \(T\) below \(T_{c}\): Either at the DW's or in the center of the domains. Available experimental data point out that the nucleation of superconductivity preferably occurs at the DW's\cite{Iavarone2014}. However it will be shown that the exact location depends on the considered type of DW's.
In addition, we show that new interesting and non-trivial features arise in the system under consideration. For example, we find an pronounced asymmetry in the \(z\)-dependence of \(\vb{j}_{\text{S}}(z)\), which occurs for both N\'{e}el DW's and \mbox{N\'{e}el--type} magnetic Sk's. This asymmetry is characterized by the in-plane dependence \(\vb*{r}_{\perp}\) of the Meissner current \(\vb{j}_{\text{S}}(\vb*{r}_{\perp})\) and stray-field \(\vb{H}_{\text{str}}(\vb*{r}_{\perp})\), which differs greatly above and below the ferromagnet and can even result in a local sign change of the Meissner current. In the absence of a superconductor, the asymmetry of the stray-field for DW's is already known\cite{mallinson_one-sided_1973}. For instance, it was recently demonstrated in experiments on artificial magnetic structures\cite{Marioni_Halbach_2018}. The asymmetry arises due to a non-zero \(\div{\vb{M}}\) term inside the ferromagnet. In the case of Bloch DW's such an asymmetry does not appear, since \(\div{\vb{M}}\) term vanishes. 

This difference between Bloch-- or \mbox{N\'{e}el--type} DW's and Sk's follows from the different orientation of the vector \(\vb{\hat{e}}_{\text{rot}}\) which describes the rotation axis of the magnetization \(\vb{M}\) along the domain wall. For instance, in the case of Bloch-- and \mbox{N\'{e}el--type} DW one can define a vector \mbox{\(\vb{\hat{e}}_{N}\equiv(\vb{\hat{e}}_{\text{rot}}\times\vb{\hat{e}}_{x})\)}, where \(\vb{\hat{e}}_{x}\) is a unit vector normal to the plane of the DW. For \mbox{N\'{e}el--type} DW's this vector is non-zero, while for \mbox{Bloch--type} DW's the vector product is zero  because the rotation vector \(\vb{\hat{e}}_{\text{rot}}\) is collinear to the vector \(\vb{\hat{e}}_{x}\). In the language of magnetic monopoles, which can be used for magnetic stray-fields, the presence of \(\vb{\hat{e}}_{N}\) translates into the existence of magnetic bulk charges. In combination with the magnetic surface charges, the stray-field components of the bulk charges results in the aforementioned asymmetry. In the case of a S/F bilayer it generally makes no sense to speak about an asymmetry, but the spatial distribution of the Meissner \(\vb{j}_{\text{S}}\) in the S still depends on the direction of the vector \(\vb{\hat{e}}_{N}\) with respect to the F film (upward or downwards).

We will begin this paper by calculating a general expression for the effective magnetic stray-field \(\vb{H}(\vb*{r})\) in an S/F/S structure generated by a nonhomogeneous two-component magnetization \(\vb{M}\), see \mbox{Sec. \ref{Section1}}. From the stray-field, we extract an expression for the Meissner current in the two superconducting region, which is then applied to describe induced currents in the presence of isolated N\'{e}el-- and \mbox{Bloch--type} Sk, see \mbox{Sec. \ref{Section2}}, as well as for various magnetic DW configurations, see \mbox{Sec. \ref{Section3}}. In \mbox{Sec. \ref{Section4}}, we use a Ginzburg-Landau model in the absence of external currents to estimate the nucleation of  superconductivity in the presence of the DW structures, we considered earlier. Note that the obtained results in this sections are independent of the type of S, as we are working with unscreened magnetic stray-fields. The universal expressions for these unscreened fields can be easily extracted from our results in the previous section. We end this work with a conclusion in \mbox{Sec. \ref{Section5}}.

\section{Stray field and Meissner current \label{Section1}}
	
We consider an S/F/S structure, that is, a ferromagnetic film of thickness \(2d_{\text{F}}\) interfaced by two superconductors at \mbox{\(z=\pm d_{\text{F}}\)}. The magnetization \(\vb{M}(\vb*{r})\) inside the ferromagnet can be written in the form
\begin{equation}
	\vb{M}(\vb*{r})=M_{0}\vb{n}(\vb*{r})
\end{equation}
where the unit vector \(\vb{n}(\vb*{r})\) is a function of the position vector \mbox{\(\vb*{r}=(\vb*{r}_{\perp},z)\)} with \(\vb*{r}_{\perp}\) lying in the \((x,y)\)-plane. In the following the magnetization is assumed to be independent of the \(z\)-coordinate .
	
We will now begin with determining the spatial distribution of the screened stray-field in the superconducting regions. The superconducting order parameters (OP) are assumed to be homogeneous. Any magnetic field  \mbox{\(\vb{H}_{\text{S}}(\vb*{r})=\qty(\vb{H}_{\text{S}}^{\perp}(\vb*{r}),\vb{H}_{\text{S}}^{z}(\vb*{r}))\)} inside the S must then satisfy the London equation which we write for the Fourier component \mbox{\(\vb{H}_{\text{S}}(\vb*{k},z)=\int\dd{\vb*{r}_{\perp}}\vb{H}_{\text{S}}(\vb*{r}_{\perp},z)\exp(-i\vb*{k}\vdot\vb*{r}_{\perp})\)}
\begin{equation}
	\partial_{zz}^{2}\vb{H}_{\text{S}}(\vb*{k},z)-\kappa^{2}\vb{H}_{\text{S}}(\vb*{k},z)=0, \quad \text{S regions}\label{Eq: London Fourier}
\end{equation}
where \mbox{\(\kappa^{2}=\abs{k}^2+\lambda_{\text{L}}^{-2}\)} and \(\lambda_{\text{L}}\) is the London penetration depth. In the general case, the two superconductors may have different London penetration depth \(\lambda_{\text{L}}^{+}\) and \(\lambda_{\text{L}}^{-}\).
The solution of Eq.(\ref{Eq: London Fourier}) is given by 
\begin{align}
	\vb{H}_{\text{S}}(\vb*{k},z)=\vb{C}_{\pm}(\vb*{k})e^{-\kappa_{\pm}\abs{z}}, \quad \text{S regions}
\end{align}
where the index \(\pm\) of the constant \mbox{\(\vb{C}_{\pm}=(\vb{C}_{\pm}^{\perp},\mathrm{C}_{\pm}^{z})\)} and \(\kappa^{\pm}\) indicates their values in the upper/lower superconducting regions, respectively.\\
	
The stray-field generated by the magnetization \(\vb{M}\) inside the F has to fulfill the magnetostatic condition \mbox{\(\curl{\vb{H}_{\text{F}}}=0\)} which allows us to define a magnetic scalar potential \(U\) with
\begin{equation}
	\vb{H}_{\text{F}}(\vb*{k},z)=-\qty(i\vb*{k}U(\vb*{k},z),\partial_{z}U(\vb*{k},z))\label{Eq: Relation Potential and H Field}\quad
\end{equation}
In the absence of the proximity effect (PE) the potential \(U\) is related to the magnetization \(\vb{M}\) via \mbox{\(\div{\vb{H}_{\text{F}}}=-4\pi\div{\vb{M}}\)} so that we can write
\begin{equation}
	\partial_{zz}^{2}U(\vb*{k},z)-\abs{k}^{2}U(\vb*{k},z)=4\pi M_{0} i\vb{k}\vdot\vb{n}_{\perp}(\vb*{k}), \quad \text{F film} \label{Eq: DGL Potential}\quad
\end{equation}
Solving Eq.(\ref{Eq: DGL Potential}) for \(U(\vb*{k},z)\) we obtain
	\begin{align}
		U(\vb*{k},z)=&4\pi M_{0}\Bigg\{A(\vb*{k})\sinh(\abs{k}z)+B(\vb*{k})\cosh(\abs{k}z)\notag\\
		&-\frac{i\vb*{k}\vdot\vb{n}_{\perp}(\vb*{k})}{\abs{k}^2}+C_{0}\delta(\vb*{k})+\frac{i\vb*{k}\vdot \vb{C}_{\vb*{r}_{\perp}}}{\abs{k}^{2}}\delta(\vb*{k})\Bigg\}
	\end{align}
with the Dirac $\delta$-function \(\delta(\vb*{k})\). The last two terms are contributions to the homogeneous solutions of Eq.(\ref{Eq: DGL Potential}). In the coordinate representation it has the form: \mbox{\(C_{0}+\vb{C}_{\perp}\vb*{r}_{\perp}\)} Note, the constant \(C_{0}\) does not affect any physical quantity, so that we can set \mbox{\(C_{0}=0\)}. The constant \(\vb{C}_{\perp}\) on the other hand, is related to a non-compensated magnetic moment \(\vb{M}_{un}\) in the F which turns to zero for \mbox{\(\vb{M}_{un}=0\)}. Using Eq.(\ref{Eq: Relation Potential and H Field}) we can determine the stray-field \(\vb{H}_{\text{F}}\) in the F film 
\begin{align}
		\vb{H}_{\text{F}}^{\perp}(\vb*{k},z)=&-4\pi M_{0} i\vb*{k} \Bigg\{A(\vb*{k})\sinh(\abs{k}z)+B(\vb*{k})\cosh(\abs{k}z)\notag\\
		&-\frac{i\vb*{k}\vdot\vb{\bar{n}}_{\perp}(\vb*{k})}{\abs{k}^2}\Bigg\}\\
		\mathrm{H}_{\text{F}}^{z}(\vb*{k},z)=&-4\pi M_{0} \abs{k}\qty{A(\vb*{k})\cosh(\abs{k}z)+B(\vb*{k})\sinh(\abs{k}z)}		
\end{align}
where we defined
\begin{equation}
		\vb{\bar{n}}_{\perp}(\vb*{k}):=\qty(\vb{n}_{\perp}(\vb*{k})-\vb{C}_{\vb*{r}_{\perp}}\delta(\vb{k}))
	\end{equation}
The constants \(A(\vb*{k})\), \(B(\vb*{k})\) and \(\vb{C}_{\pm}(\vb*{k})\) can be found using the matching conditions for the magnetic field and the magnetic induction at the S/F interfaces. They are reduced to the continuity of the tangential components of the in-plane field \(\vb{H}^{\perp}(\vb*{k},z)\) and the normal component of the magnetic induction \mbox{\(\mathrm{B}^{z}(\vb*{k},z)=\mathrm{H}^{z}(\vb*{k},z)+4\pi M_{0}\mathrm{n}_{z}(\vb*{k})\)}, \textit{i.e.},
	\begin{align}
		\vb{H}_{\text{S}}^{\perp}(\pm d_{\text{F}})=&\vb{H}_{\text{F}}^{\perp}(\pm d_{\text{F}})\label{Eq: Interface Cond1}\\
		\mathrm{H}_{\text{S}}^{z}(\pm d_{\text{F}})=&\mathrm{H}_{\text{F}}^{z}(\pm d_{\text{F}})+4\pi M_{0}\mathrm{n}_{z}(\vb*{k})\label{Eq: Interface Cond2}
	\end{align}
In addition, the in-plane component of \(\vb{H}_{\text{S}}\) is coupled to the normal component via the equation \mbox{\(\div{\vb{H}_{\text{S}}}=0\)} so that 
\begin{equation}
		\vb{C}_{\pm}^{\perp}(\vb*{k})=\mp\frac{i\vb*{k}}{\abs{k}}\frac{\kappa_{\pm}}{\abs{k}}\mathrm{C}_{\pm}^{z}(\vb*{k}) \label{Eq: Divergenzless HS}
\end{equation}
	\begin{widetext}
Using Eqs.(\ref{Eq: Interface Cond1}-\ref{Eq: Divergenzless HS}) we can determine the coefficients \(A(\vb*{k})\) and \(B(\vb*{k})\), which are given by 
	\begin{align}
		A(\vb*{k})=&\mathrm{n}_{z}(\vb*{k})\frac{\kappa_{+}D_{2}^{-}(k)+\kappa_{-}D_{2}^{+}(k)}{\abs{k}D(k)}+\vb{\bar{n}}_{\perp}(\vb*{k})\frac{i\vb*{k}}{\abs{k}}\frac{D_{2}^{-}(k)-D_{2}^{+}(k)}{D(k)}\\
		B(\vb*{k})=&\mathrm{n}_{z}(\vb*{k})\frac{\kappa_{+}D_{1}^{-}(k)-\kappa_{-}D_{1}^{+}(k)}{\abs{k}D(k)}+\vb{\bar{n}}_{\perp}(\vb*{k})\frac{i\vb*{k}}{\abs{k}}\frac{D_{1}^{-}(k)+D_{1}^{+}(k)}{D(k)}
	\end{align}
with \(D(k)=D_{1}^{-}(k)D_{2}^{+}(k)+D_{1}^{+}(k)D_{2}^{-}(k)\), where
\begin{align}
		D_{1}^{\pm}(k)=&\abs{k}\sinh(\abs{k}d_{\text{F}})+\kappa_{\pm}\cosh(\abs{k}d_{\text{F}})\\
		D_{2}^{\pm}(k)=&\abs{k}\cosh(\abs{k}d_{\text{F}})+\kappa_{\pm}\sinh(\abs{k}d_{\text{F}})
\end{align}
The coefficient \(\vb{C}_{\pm}(\vb*{k})\) is given by 
	\begin{align}
		C_{\pm}^{z}(k)=-4\pi M_{0}\abs{k}\sinh(\abs{k}d_{\text{F}})e^{\kappa_{\pm} d_{\text{F}}}&\left[\pm\frac{i \vb*{k}}{\abs{k}}\frac{\vb{\bar{n}}_{\perp}(\vb*{k})}{D(k)}\qty(\qty{D_{1}^{-}(k)+D_{1}^{+}(k)}\pm\qty{\kappa_{-}-\kappa_{+}}\cosh(\abs{k}d_{\text{F}}))\right.\notag\\
		&\left.-\frac{\mathrm{n}_{z}(\vb*{k})}{D(k)}\qty(\qty{D_{2}^{-}(k)+D_{2}^{+}(k)}\pm\qty{\kappa_{-}-\kappa_{+}}\sinh(\abs{k}d_{\text{F}}))\right]
	\end{align}
For the sake of simplicity, we will from now on consider two identical superconducting materials, \textit{i.e.}, \(\lambda_{\text{L}}^{+}=\lambda_{\text{L}}^{-}\). In this case the expression for the coefficients can be reduced to 
\begin{align}
		A(\vb*{k})=\frac{\kappa}{\abs{k}}\frac{\mathrm{n}_{z}(\vb*{k})}{D_{1}(k)},
		&&B(\vb*{k})=\frac{i\vb*{k}}{\abs{k}}\frac{\vb{\bar{n}}_{\perp}(\vb*{k})}{D_{2}(k)}
	\end{align}
and 
\begin{equation}
		\mathrm{C}_{\pm}^{z}(\vb*{k})=-4\pi M_{0}\abs{k}\sinh(\abs{k}d_{\text{F}})e^{\kappa d_{\text{F}}}\qty[\pm\frac{i\vb*{k}}{\abs{k}}\frac{\vb{\bar{n}}_{\perp}(\vb*{k})}{D_{2}(k)}-\frac{\mathrm{n}_{z}(\vb*{k})}{D_{1}(k)}]
\end{equation}
With this we obtain the \(k\)-space representation of the screened stray-field in an S/F/S junction for two identical superconductors.

\noindent In the S region \(\abs{z}>d_{\text{F}}\):
	\begin{align}
		\vb{\tilde{H}}_{\perp}^{(\text{S})}(\vb*{k},z)=&\kappa\sinh(\abs{k}d_{\text{F}})\frac{i\vb*{k}}{\abs{k}}\qty[\frac{i\vb*{k}}{\abs{k}}\frac{\vb{\bar{n}}_{\perp}(\vb*{k})}{D_{2}(k)}\mp\frac{\mathrm{n}_{z}(\vb*{k})}{D_{1}(k)}]e^{-\kappa\abs{z\mp d_{\text{F}}}}\label{Eq: HSperp kspace}\\
		\mathrm{\tilde{H}}_{z}^{(\text{S})}(\vb*{k},z)=&-\abs{k}\sinh(\abs{k}d_{\text{F}})\qty[\pm\frac{i\vb*{k}}{\abs{k}}\frac{\vb{\bar{n}}_{\perp}(\vb*{k})}{D_{2}(k)}-\frac{\mathrm{n}_{z}(\vb*{k})}{D_{1}(k)}]e^{-\kappa\abs{z\mp d_{\text{F}}}}\label{Eq: Hz kspace}
	\end{align}
	\noindent In the F film \(\abs{z}<d_{\text{F}}\):
	\begin{align}
		\vb{\tilde{H}}_{\perp}^{(\text{F})}(\vb*{k},z)=&-\frac{i\vb*{k}}{\abs{k}}\qty[\frac{i\vb*{k}}{\abs{k}}\vb{\bar{n}}_{\perp}(\vb*{k})\qty(\frac{\abs{k}\cosh(\abs{k}z)}{D_{2}(k)}-1)+\frac{\mathrm{n}_{z}(\vb*{k})\kappa\sinh(\abs{k}z)}{D_{1}(k)}]\label{Eq: HFperp kspace}\\
		\mathrm{\tilde{H}}_{z}^{(\text{F})}(\vb*{k},z)=&-
		\qty[\frac{i\vb*{k}}{\abs{k}}\frac{\vb{\bar{n}}_{\perp}(\vb*{k})\abs{k}\sinh(\abs{k}z)}{D_{2}(k)}+\frac{\mathrm{n}_{z}(\vb*{k})\kappa\cosh(\abs{k}z)}{D_{1}(k)}] \label{Eq: HFz kspace}
	\end{align}
where we expressed the results in terms of a dimensionless field \(\tilde{\vb{H}}=\vb{H}/4\pi M_{0}\).
\end{widetext}

The obtained expressions describe the screened stray-field in an S/F/S structure. By taking \mbox{\(\lambda_{\text{L}}\rightarrow\infty\)}, \textit{i.e.}, \mbox{\(\kappa\rightarrow k\)}, we can also extract the unscreened stray-field which would be present in the absence of superconductors. In this limit, the result describes the general distribution of the stray-field created by a nonhomogeneous magnetization in a F in vacuum. The associated vector potential is later used to estimate the nucleation of superconductivity. The origin of the screening field that leads to the effective stray-field in Eq.(\ref{Eq: HFperp kspace},\ref{Eq: HFz kspace}), are the induced supercurrents inside the S. The supercurrent (Meissner current) \(\vb{j}_{\text{S}}\) can be determined using Amp\`{e}re's law \mbox{\(\curl{\vb{H}}=\frac{4\pi}{c}\vb{j}_{\text{\text{S}}}\)}.
\begin{equation}
		\vb{j}_{\text{S}}(\vb*{r}_{\perp},z)=\frac{c}{4\pi}\qty[\qty(\nabla_{\vb*{r}_{\perp}}+\hat{\vb{e}}_{z}\partial_{z})\times\qty(\vb{H}_{\perp}(\vb*{r}_{\perp},z)+\hat{\vb{e}}_{z}\mathrm{H}_{z}(\vb*{r}_{\perp},z))]
\end{equation}
In the Fourier representation we further obtain
\begin{equation}
		\vb{j}_{\text{S}}(\vb*{k},z)=\frac{c}{4\pi}\qty[\qty(i\vb*{k}+\hat{\vb{e}}_{z}\partial_{z})\times\qty(\vb{H}_{\perp}(\vb*{k},z)+\hat{\vb{e}}_{z}\mathrm{H}_{z}(\vb*{k},z))]
	\end{equation}
It can easily be shown, that the supercurrent disappears within the F, which is the expected result when the PE is absent.  Outside the ferromagnet \mbox{\(\abs{z}>d_{\text{F}}\)}, we obtain the following expression
	\begin{equation}
		\vb{j}_{\text{S}}(\vb*{k},z)=\frac{cM_{0}\lambda_{\text{L}}^{-2}}{\abs{k}}\qty(\hat{\vb{e}}_{z}\times\frac{i\vb*{k}}{\abs{k}})\mathrm{\tilde{H}}_{z}^{(\text{S})}(\vb*{k},z)\label{Eq:MeissnerCurrent}
	\end{equation}
from which we can directly derive the vector potential \(\vb{A}\) in the superconductor using \mbox{\(\vb{j}_{\text{S}}=-c\lambda_{\text{L}}^{-2}\vb{A}^{(\text{S})}/4\pi\)}
\begin{equation}
		\vb{A}^{(\text{S})}(\vb*{k},z)=-\frac{4\pi M_{0}}{\abs{k}}\qty(\hat{\vb{e}}_{z}\times\frac{i\vb*{k}}{\abs{k}})\mathrm{\tilde{H}}_{z}^{(\text{S})}(\vb*{k},z) \label{Eq: Vectorpotential S}
\end{equation}
	\begin{figure*}[tbp]
		\includegraphics[width=\linewidth]{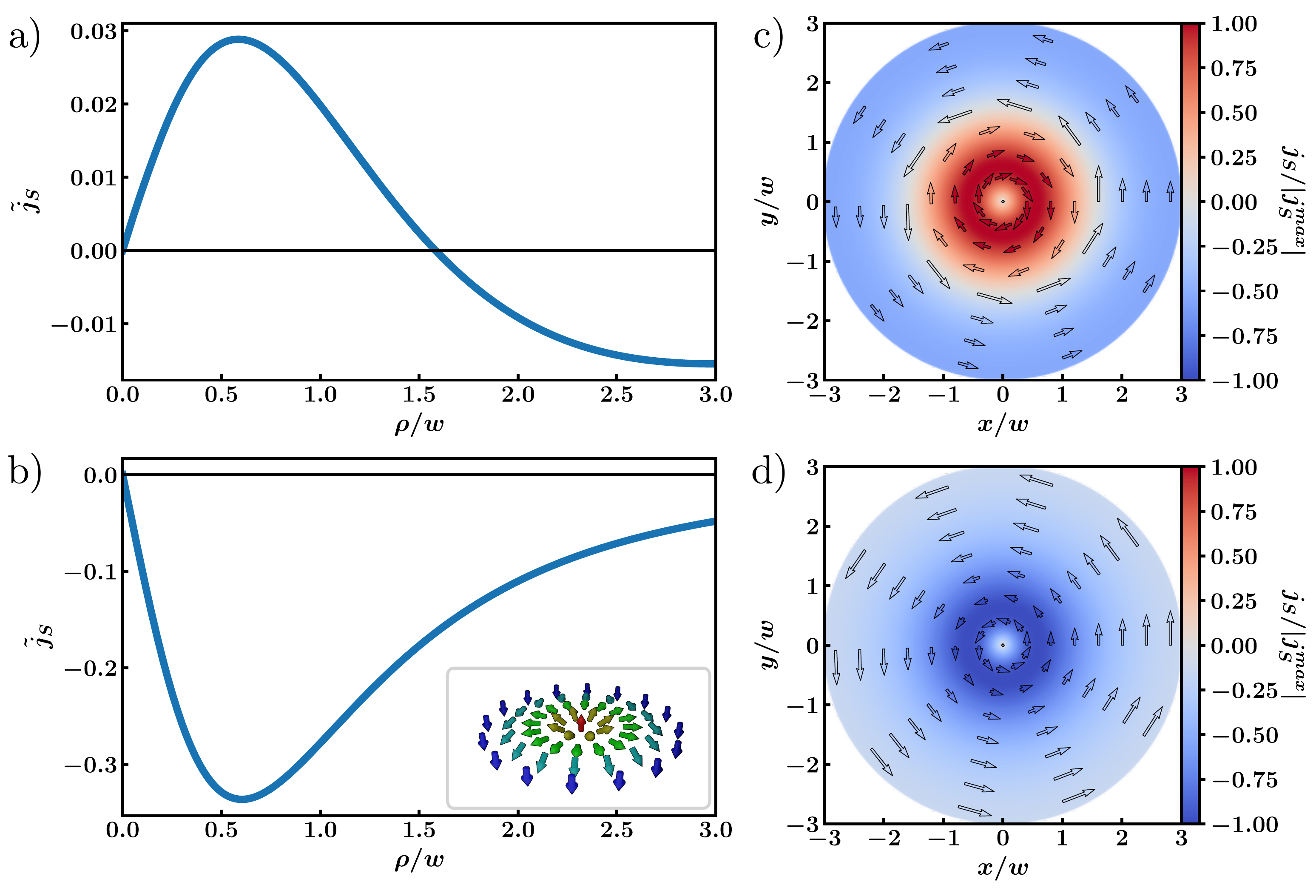}
		\caption{(Color online) Radial plots of the normalized Meissner current \mbox{\(\tilde{j}_{\text{S}}=j_{\text{S}}/cM_{0}\lambda_{\text{S}}^{-2}w\)} generated by the stray-field of a \mbox{N\'{e}el Sk}, shown in the inset of (b), in the a) upper and  b) lower superconducting regions for \(\lambda_{\text{L}}=w\) and \(d_{\text{F}}=10w\). The arrows in the corresponding 2D plots c) and d) indicate the direction of the circular screening current. The color map is normalized w.r.t. to the maximal value of \(j_{\text{S}}\) in the upper c) and lower d) superconductors, respectively. The asymmetry of the stray-fields in the upper and lower superconductors leads to a sign change of the Meissner current in the upper S above the Sk region. \label{Fig: Meissner Neel}}
	\end{figure*}
	
\section{Isolated Skyrmion \label{Section2}}
In this section, we will set the magnetization profile \(\vb{M}\) to describe an isolated magnetic skyrmion (Sk) in a ferromagnetic background. It is assumed that the Sk's are stabilized by an underlying chiral interaction resulting in either Bloch-- or \mbox{N\'{e}el--type} Sk's. The magnetization profile has a cylindrical symmetry and varies along the radial direction \(\vb*{\rho}\) so that \mbox{\(\vb*{r}_{\perp}=\vb*{\rho}\)}.
The unit vector \(\vb{n}\) of a chiral Bloch or N\'{e}el Sk can then be written as
\begin{align}
		\vb{n}_{\perp}(\vb*{\rho})=&\frac{\vb*{\rho}}{\rho}\sin(\theta(\rho))\Theta(r_{\text{Sk}}-\rho), \quad &&\text{N\'{e}el Sk} \label{Eq: M inplane Neel}\\
		\vb{n}_{\perp}(\vb*{\rho})=&\frac{\vb{\hat{e}}_{z}\cross\vb*{\rho}}{\rho}\sin(\theta(\rho))\Theta(r_{\text{Sk}}-\rho), \quad &&\text{Bloch Sk} \label{Eq: M inplane Bloch}
	\end{align}
for the in-plane component and
	\begin{equation}
		\mathrm{n}_{z}(\vb*{\rho})=\qty[1+\cos(\theta(\rho))]\Theta(r_{\text{Sk}}-\rho)-1 , \quad \text{Bloch \& N\'{eel} Sk} \label{Eq: M out-ofplane}
	\end{equation}
for the out-of-plane component. Here, \(\theta(\rho)\) describes the angular variation of the magnetization w.r.t. the \(z\)-axis and \(\Theta(r_{\text{Sk}}-\rho)\) is a Heaviside step function with \(r_{\text{Sk}}\) being the skyrmion radius. The Fourier components of \(\vb{n}(\vb*{\rho})\) are equal to 
\begin{align}
		\vb{n}_{\perp}(\vb*{k})=&-2\pi\frac{i\vb*{k}}{k} m_{\perp}(k), \quad &&\text{N\'{e}el Sk}\\
		\vb{n}_{\perp}(\vb*{k})=&-2\pi\frac{\vb{\hat{e}}_{z}\cross\vb*{k}}{k} m_{\perp}(k), \quad &&\text{Bloch Sk}
\end{align}	
and 
\begin{equation}
		\mathrm{n}_{z}(\vb*{k})=2\pi\qty[m_{z}(k)-2\pi\delta(\vb*{k})], \quad \text{Bloch \& N\'{eel} Sk}
\end{equation}
The functions \(m_{\perp}(k)\) and \(m_{z}(k)\) are defined as
	\begin{align}
		m_{\perp}(k)=&\int_{0}^{r_{\text{Sk}}}\dd{\rho}\rho J_{1}(k\rho)\sin(\theta(\rho))\\
		m_{z}(k)=&\int_{0}^{r_{\text{Sk}}}\dd{\rho}\rho J_{0}(k\rho)\qty[1+\cos(\theta(\rho))]
	\end{align}
where \(J_{n}(x)\) is the Bessel-function of the first kind of order \(n\). The angular dependence of the Sk profile \(\theta(\rho)\), is described using a circular  \(360^{\circ}\)--domain wall Ansatz\cite{Romming_field_2015}.
	\begin{equation}
		\theta(\rho)=\sum_{\pm}\arcsin(\tanh(-\frac{\rho\mp c}{w/2}))\label{Eq: Magnetizationprofile}
	\end{equation}
with \(c\) being the size of the domain core and \(w\) is the domain wall width. For the remainder of this work, we set \(c=0\,\mathrm{nm}\). Using Eq.(\ref{Eq: Magnetizationprofile}), one can estimate the radius \(r_{\text{Sk}}\) of the Sk. It should be noted that the expressions in this section can be used for any radially symmetric magnetization profile. 
		
Using the obtained result from the previous section, we will begin analyzing the effective stray-field generated by a Sk in our S/F/S structure in the case of two identical superconductors. Afterwards we will determine the corresponding induced Meissner currents. Taking into account that for a Bloch Sk \mbox{\(\vb{n}_{\perp}\propto (\vb{\hat{e}}_{z}\cross\vb*{k})\)} (see Eq.(\ref{Eq: M inplane Bloch})), we see that the first term in Eqs.(\ref{Eq: HSperp kspace}-\ref{Eq: HFz kspace}) vanishes. This means that the individual components of the stray-field are either symmetric \mbox{\(\mathrm{H}_{z}(z)=\mathrm{H}_{z}(-z)\)}  or anti-symmetric functions \mbox{\(\vb{H}_{\perp}(z)=-\vb{H}_{\perp}(-z)\)} of \(z\). On the other hand, the in-plane magnetization of a N\'{e}el Sk \mbox{\(\vb{n}_{\perp}\propto \vb*{k}\)}. Hence, in this case \(\vb{H}(z)\neq\vb{H}(-z)\) which describes an asymmetry of the magnetic stray-field. This asymmetry is a typical feature of stray-fields  generated by magnetic textures with N\'{e}el--like magnetization\cite{Mallinson_Onesided_1973,Marioni_Halbach_2018}.

In order to fully determine the magnetic stray-field and the Meissner current, we first need to specify the value of the constant \(\vb{C}_{\vb*{r}_{\perp}}\). Using the condition that the spatial average of the in-plane component of the stray-field vanishes, \textit{i.e.}, \mbox{\(\int\dd{\vb*{r}}\vb{H}_{\perp}(\vb*{\rho},z)=0\)}, we get an additional equation for \(\vb{C}_{\vb*{r}_{\perp}}\). For the case of an isolated Sk this constant is equal to zero \mbox{\(\vb{C}_{\vb*{r}_{\perp}}=0\)}. The real-space representation of the screened stray-field in Eqs.(\ref{Eq: HSperp kspace}-\ref{Eq: HFz kspace}) can now be easily expressed as

\begin{widetext}
	\begin{figure*}[tbp]
		\includegraphics[width=\linewidth]{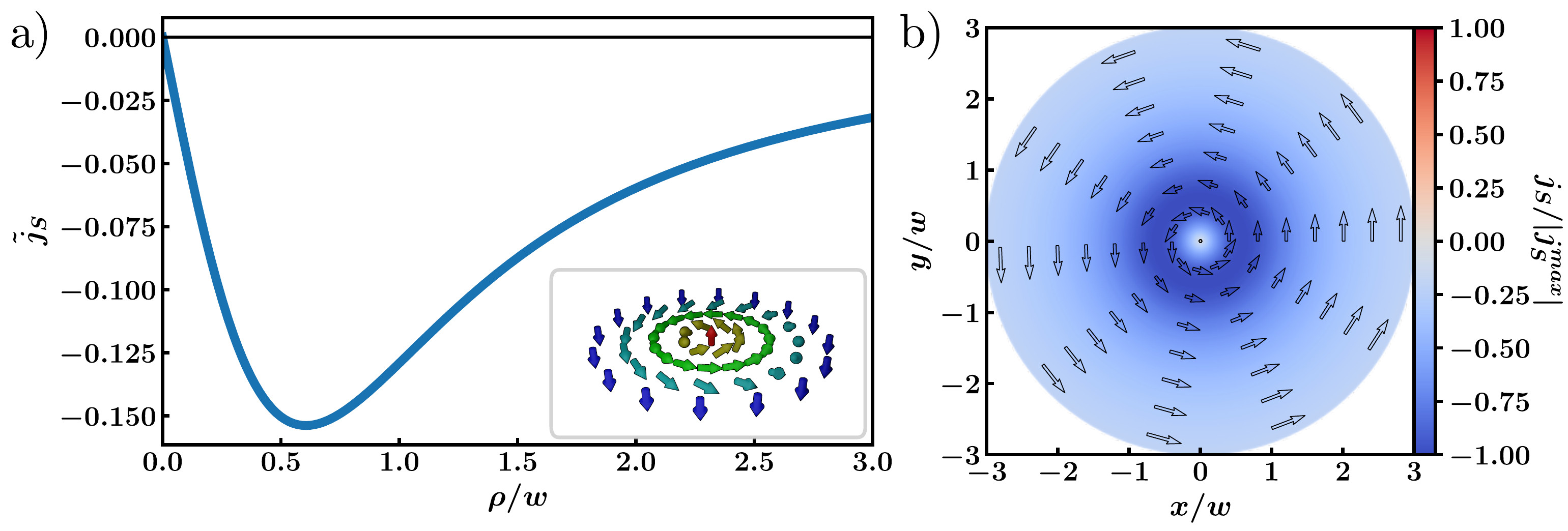}
		\caption{(Color online) a) Radial plots of the normalized Meissner current \mbox{\(\tilde{j}_{\text{S}}=j_{\text{S}}/cM_{0}\lambda_{\text{S}}^{-2}w\)} generated by the stray-field of a Bloch Sk (shown in the inset) for \(\lambda_{\text{L}}=w\) and \(d_{\text{F}}=10w\). b) 2D plot of the Meissner current with arrows indicating the direction of the circular screening current. The color map is normalized w.r.t. to the maximal value of \(j_{\text{S}}\). The current is identical for both superconducting regions. \label{Fig: Meissner Bloch}}
	\end{figure*}
\noindent In the S region \(\abs{z}>d_{\text{F}}\):
	\begin{align}
		\vb{\tilde{H}}_{\perp}^{(\text{S})}(\vb*{\rho},z)=&-\frac{\vb*{\rho}}{\rho}\int_{0}^{\infty}\dd{k}kJ_{1}(k\rho)\kappa\sinh(kd_{\text{F}})\qty[\frac{\bar{m}_{\perp}(k)}{D_{2}(k)}\mp\frac{m_{z}(k)}{D_{1}(k)}]e^{-\kappa\abs{z\mp d_{\text{F}}}}\\
		\mathrm{\tilde{H}}_{z}^{(\text{S})}(\vb*{\rho},z)=&-\int_{0}^{\infty}\dd{k}kJ_{0}(k\rho)k\sinh(kd_{\text{F}})\qty[\pm\frac{\bar{m}_{\perp}(k)}{D_{2}(k)}-\frac{m_{z}(k)}{D_{1}(k)}]e^{-\kappa\abs{z\mp d_{\text{F}}}}
	\end{align}
In the F film \mbox{\(\abs{z}<d_{\text{F}}\)}:
\begin{align}
		\vb{\tilde{H}}_{\perp}^{(\text{F})}(\vb*{\rho},z)=&\frac{\vb*{\rho}}{\rho}\int_{0}^{\infty}\dd{k}kJ_{1}(k\rho)\qty[\bar{m}_{\perp}(k)\qty(\frac{k\cosh(kz)}{D_{2}(k)}-1)+m_{z}(k)\frac{\kappa\sinh(kz)}{D_{1}(k)}]\\
		\mathrm{\tilde{H}}_{z}^{(\text{F})}(\vb*{\rho},z)=&-\int_{0}^{\infty}\dd{k}kJ_{0}(k\rho)\qty[\bar{m}_{\perp}(k)\frac{k\sinh(kz)}{D_{2}(k)}+m_{z}(k)\frac{\kappa\cosh(kz)}{D_{1}(k)}]+\frac{1}{2}
\end{align}
\end{widetext}

where we inserted the magnetization profile Eq.(\ref{Eq: M inplane Neel}-\ref{Eq: M out-ofplane}) and defined 
\begin{equation}
		\bar{m}_{\perp}(k)=\begin{cases}
		m_{\perp}(k) , &\quad \text{N\'{e}el Sk}\\
		0,  &\quad \text{Bloch Sk}
		\end{cases}
\end{equation}	
Analogously, the Meissner current can be found using Eq.(\ref{Eq:MeissnerCurrent}), which has the following form in real-space
	\begin{align}
		\vb{j}_{\text{S}}(\rho,z)=&cM_{0}\lambda_{\text{L}}^{-2}\int_{0}^{\infty}\dd{k}kJ_{1}(k\rho)\sinh(kd_{\text{F}})\notag\\
		&\times\qty[\pm\frac{\bar{m}_{\perp}(k)}{D_{2}(k)}-\frac{m_{z}(k)}{D_{1}(k)}]e^{-\kappa\abs{z\mp d_{\text{F}}}}\vb{\hat{e}}_{\varphi}\label{Eq: SK Meissner current}
	\end{align}
The stray-field induces circulating supercurrents pointing in \(\vb{\hat{e}}_{\varphi}\)-direction. Since the supercurrent is linked to the stray-field, the Meissner current also features the asymmetry which is related to the magnetization profile of the N\'{e}el Sk. Using Eq.(\ref{Eq: SK Meissner current}), this asymmetry can be identified by the changing sign in the term associated with the in-plane contribution of the magnetization. In \mbox{Fig. \ref{Fig: Meissner Neel}} we show the spatial dependence of the Meissner current \(\vb{j}_{\text{S}}(\rho,\pm d_{\text{F}})\) in the upper (a) and c)) and the lower (b) and d)) superconductors in the presence of a \mbox{N\'{e}el--type} Sk in the ferromagnetic material. The curves are displayed for the parameters \mbox{\(\lambda_{\text{F}}=w\)} and \mbox{\(d_{\text{F}}=10w\)}. As expected, we observe a strong asymmetry in the dependence \(\vb{j}_{\text{S}}(\rho, d_{\text{F}})\) and \(\vb{j}_{\text{S}}(\rho,- d_{\text{F}})\) in the upper and lower superconductors. The current \(\vb{j}_{\text{S}}(\rho,d_{\text{F}})\) in the upper S changes sign at some finite distance within the Sk region whereas the current \(\vb{j}_{\text{S}}(\rho,- d_{\text{F}})\) remains negative for all \(\rho\). Note that the sign reversal of the Meissner current in S/F systems has been found earlier \cite{bergeret_spin_2004,volkov_spin_2019,mironov_electromagnetic_2018}, but its underlying mechanism was different as it was related to the proximity effect. In the case of Bloch Sk, all the mentioned features are missing and the Meissner current in both superconducting regions is the same, see \mbox{Fig. \ref{Fig: Meissner Bloch}}.
\begin{figure}[t]
	\includegraphics[width=\linewidth]{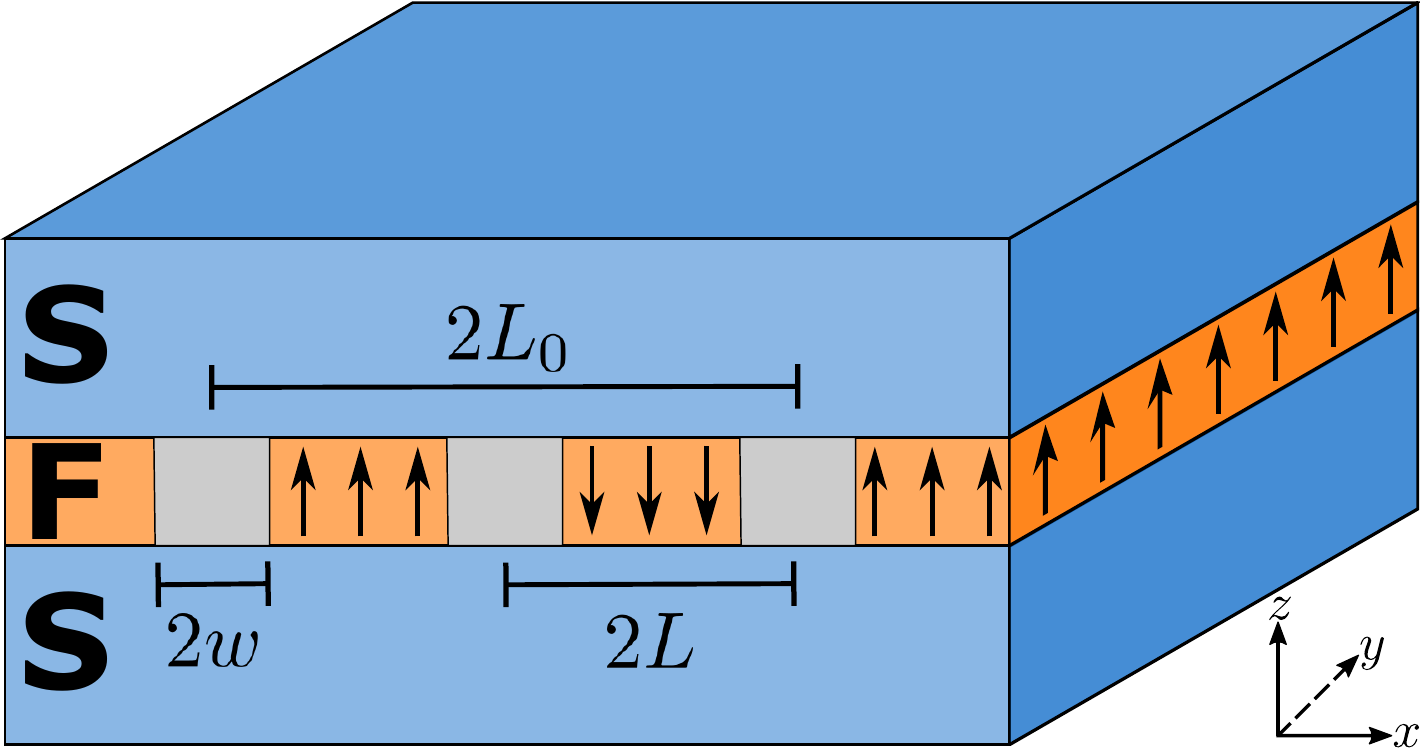}
	\caption{(Color online) Schematic picture of the S/F/S structure with DWs under consideration. Arrows describe the magnetization vectors \(\vb{M}\) in the domains with orientation chosen along the \(z\)-axis. The grey colour indicate the domain-wall regions. \label{Fig: SFS_domain}}
\end{figure}
	\begin{figure*}[thp]
		\includegraphics[width=\linewidth]{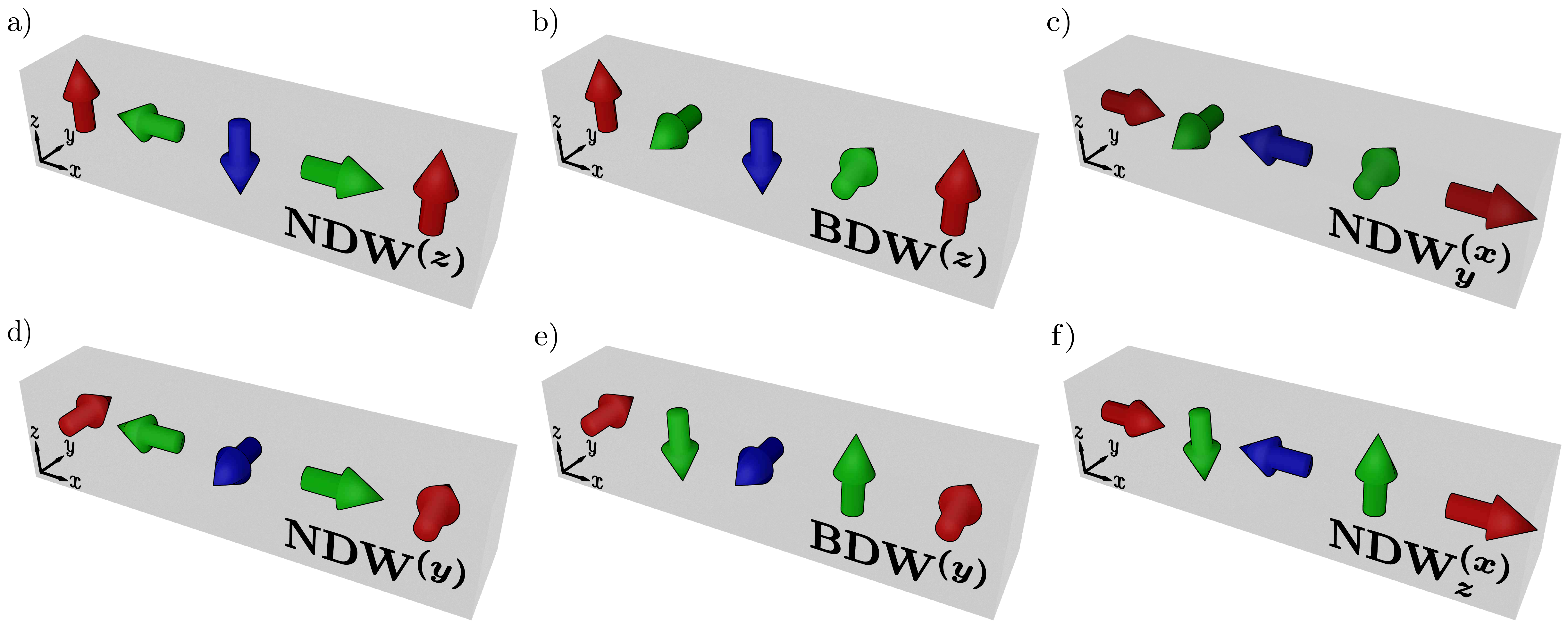}
		\caption{(Color online) Schematic representation of a single period of different periodic flat domain wall textures of \mbox{N\'{e}el--type} (NDW) and \mbox{Bloch--type} (BDW). The magnetization continuously changes from the domain in the center (blue arrow) to  the outer domains (red arrows) via the domain walls (green arrows). The superscripts indicate the orientation of the magnetization across the domain. For orientations along the \(x\)-direction, there are two possible NDW configuration described by the subscripts \(y\) and \(z\). These subscripts describe the axis over which the magnetization changes along the DW. \label{Fig: Schema}}
	\end{figure*}

\section{Flat Domain Walls\label{Section3}}
	In this section we consider the magnetization profiles of several different periodic flat DW's,  as illustrated in Fig. \ref{Fig: SFS_domain}. The alignment of magnetization changes across the DWs as a function of the \(x\)-coordinate, \textit{i.e.}, \(\vb*{r}_{\perp}=x\vb{\hat{e}}_{x}\) with \(\vb{\hat{e}}_{x}\) being the corresponding unit-vector.  The period of the structures is \(2L_{0}\). This enables us to expand all function as a Fourier series: For example, the vector \(\vb{n}(x,z)\) is represented as
	\begin{equation}
		\vb{n}(x,z)=\sum_{n=-\infty}^{\infty}\vb{n}(k_{n},z)\exp(ik_{n}x)
	\end{equation}
	with
	\begin{equation}
		\vb{n}(k_{n},z)=\frac{1}{2L_{0}}\int_{-L_{0}}^{L_{0}}\dd{x}\vb{n}(x,z)\exp(-ik_{n}x)\label{Eq: FourierSeries}
	\end{equation}
	where \mbox{\(k_{n}=\pi n/L_{0}\)}. Below we drop the subindex \(n\) for brevity.
	
Now suppose that the vector \(\vb{n}(x)\) depends only on the \(x\)-coordinate, {\it i.e.} it is completely described by its  \(x\)-component \mbox{\(\vb{k}=(k,0)\)}. In this case, the expression for the normalized magnetic stray-field \mbox{\((\tilde{H}_{x}(k,z),0,\tilde{H}_{z}(k,z))\)} and the Meissner current \(j_{\text{S}}\) can be obtained in the same manner as in \mbox{Sec. \ref{Section1}}.  For instance, for two identical S, we obtain the magnetic stray-field \(\vb{H}^{(\text{S})}(k,z)\) by substituting \mbox{\(\vb{n}_{\perp}(k)\rightarrow (\mathrm{n}_{x}(k),0)\)} and \mbox{\(\vb{C}_{r_{\perp}}\rightarrow (C_{x},0)\)} in Eq.(\ref{Eq: HSperp kspace}-\ref{Eq: Hz kspace}). For periodic DW's, one further needs to replace \mbox{\(\delta(k)\rightarrow\sin(kL_{0})/kL_{0}\)}, which follows from the finite range of integration in Eq.(\ref{Eq: FourierSeries}). Finally, the normalized field components in the superconducting regions \mbox{\(\abs{z}>d_{\text{F}}\)} are:
\begin{align}
		\tilde{\mathrm{H}}_{x}^{(\text{S})}(k,z)=&-\kappa\sinh(\abs{k}d_{\text{F}})\qty[\frac{\mathrm{\bar{n}}_{x}(k)}{D_{2}(k)}\pm\frac{ik}{\abs{k}}\frac{\mathrm{n}_{z}(k)}{D_{1}(k)}]e^{-\kappa\abs{z\mp d_{\text{F}}}}\label{Eq: DW HSx}\\
		\tilde{\mathrm{H}}_{z}^{(\text{S})}(k,z)=&-\abs{k}\sinh(\abs{k}d_{\text{F}})\qty[\pm\frac{ik}{\abs{k}}\frac{\mathrm{\bar{n}}_{x}(k)}{D_{2}(k)}-\frac{\mathrm{n}_{z}(k)}{D_{1}(k)}]e^{-\kappa\abs{z\mp d_{\text{F}}}}
\end{align}
with
\begin{equation}
		\mathrm{\bar{n}}_{x}(k):=\mathrm{n}_{x}(k)-C_{x}\frac{\sin(kL_{0})}{kL_{0}}\label{Eq: barnx}
\end{equation}
and analogously within the ferromagnet \(\abs{z}<d_{\text{F}}\):
\begin{align}
		\tilde{\mathrm{H}}_{x}^{(\text{F})}(k,z)=&\qty[\mathrm{\bar{n}}_{x}(k)\qty(\frac{\abs{k}\cosh(\abs{k}z)}{D_{2}(k)}-1)-\frac{ik}{\abs{k}}\frac{\mathrm{n}_{z}(k)\kappa\sinh(\abs{k}z)}{D_{1}(k)}]\\
		\tilde{\mathrm{H}}_{z}^{(\text{F})}(k,z)=&-\qty[\frac{ik}{\abs{k}}\frac{\mathrm{\bar{n}}_{x}(k)\abs{k}\sinh(\abs{k}z)}{D_{2}(k)}+\frac{\mathrm{n}_{z}(k)\kappa\cosh(\abs{k}z)}{D_{1}(k)}]
\end{align}	
The Meissner current can be extracted from Eq.(\ref{Eq:MeissnerCurrent}). The supercurrent flows in \(y\)-direction \mbox{\(\vb{j}_{\text{S}}(k,z)=(0,j_{\text{S}}(k,z),0)\)} and has the magnitude
\begin{equation} 	   	    	j_{\text{S}}(k,z)=cM_{0}\lambda_{\text{L}}^{-2}\sinh(\abs{k}d_{\text{F}})\qty[\pm\frac{\mathrm{\bar{n}}_{x}(k)}{D_{2}(k)}+\frac{ik}{\abs{k}}\frac{\mathrm{n}_{z}(k)}{D_{1}(k)}]e^{-\kappa\abs{z\mp d_{\text{F}}}}\label{Eq: DW jS}
\end{equation}    
where once again the \(\pm\) indicates the solution in the upper or lower S region, respectively.
The current in coordinate representation \(\vb{j}(x,z)\) can be calculated using
    \begin{equation}
    	\vb{j}(x,z)=\sum_{k}\vb{j}(k,z)\exp(ikx)
    \end{equation}
Having determined the expressions for \(\vb{H}_{str}\) and \(\vb{j}_{\text{S}}\) for an arbitrary type of DW, we need to specify the precise magnetic texture. Its components can be expressed in terms of the function \(\mathrm{n}_{even}\) and \(\mathrm{n}_{odd}\) which are characterized by an even or odd dependency on \(x\) or \(k\). Since we are interested in a qualitative spatial dependence of all quantities (the fields and the Meisner currents), we approximate \(\mathrm{n}_{even,odd}\) by
\begin{widetext}
	\begin{align}
		n_{odd}(x)=&\cos(\frac{\pi}{2}\frac{(x-L)}{w})\theta(w-\abs{x-L})-\cos(\frac{\pi}{2}\frac{(x+L)}{w})\theta(w-\abs{x+L})\label{Eq:noddx}\\
		n_{even}(x)=&\qty[1-\sin(\frac{\pi}{2}\frac{(x-L)}{w})]\theta(w-\abs{x-L})+\qty[1+\sin(\frac{\pi}{2}\frac{(x+L)}{w})]\theta(w-\abs{x+L})+2\theta(L-w-\abs{x})-1
	\end{align}
with the domain wall width \(2w\) and the size of the domain \mbox{\(2(L-w)\)}. That is, we assume that the rotation angle of the vector \(\vb{n}\) outside the DW's \mbox{(\(\abs{x\mp L}>w\))} remains constant whereas it changes linearly inside the DW's \mbox{(\(\abs{x\mp L}<w\))}. This approximation allows us to present results in a simple analytical form. Outside the interval \mbox{\(\abs{x}<L_{0}\)}, \(n(x)\) is a periodic function of \(x\): \mbox{\(n(x)=n(x+2L_{0})\)}. The Fourier components of \(n_{odd}(x)\) and \(n_{even}(x)\) are equal to
	\begin{align}
		n_{odd}(k)=&\frac{2\pi iw}{2L_{0}}\frac{\cos(kw)\sin(kL)}{k^2w^2-\qty(\pi/2)^{2}}=\frac{2\pi iw}{2L_{0}} f(k)\label{Eq:noddk}\\
		n_{even}(k)=&\frac{\pi^{2}}{2L_{0}k}\frac{\cos(kw)\sin(kL)}{k^2w^2-\qty(\pi/2)^{2}}-\frac{2\sin(kL_{0})}{2L_{0}k}=\frac{2\pi w}{2L_{0}} F(k)\label{Eq:nevenk}
	\end{align}
	with 
	\begin{align}
		f(k)=&\frac{\cos(kw)\sin(kL)}{k^2w^2-\qty(\pi/2)^{2}}\label{Eq:foddk}\\
		F(k)=&\frac{\pi}{2kw}f(k)-\frac{\sin(kL_{0})}{\pi kw}\label{Eq:fevenk}
	\end{align}
Obviously, \(f(k)\) is also an odd function, whereas \(F(k)\) is an even function of \(k\). It should be noted that the limiting case of the DW width \(w=0\) was analyzed in Ref.\cite{Laiho2003,stankiewicz_magnetic_1997,bulaevskii_ferromagnetic_2000,Sonin_ferromagnetic_2002}.

In our model, the vector \(\vb{n}\) has two non-zero components that allow the construction of six different magnetic textures (see \mbox{Fig. \ref{Fig: Schema}}). They are characterized by vectors \(\vb{n}\) with the following components: \((0,n_{even},n_{odd})\), \((0,n_{odd},n_{even})\), \((n_{even},0,n_{odd})\), \((n_{odd},0,n_{even})\) and \((n_{even},n_{odd},0)\), \((n_{odd},n_{even},0)\). Note that we are working with the underlying assumption of fixed chirality, \textit{i.e}, the vector \(\vb{n}\) rotates in the same direction within the DW's, which is either clock-wise or counter-clockwise. Another chirality may be obtained if the rotation of the vector \(\vb{n}\) in adjacent DW's occurs in different directions; then the function \(n_{odd}\) should be replaced by \mbox{\(n_{odd}\Rightarrow\tilde{n}_{even}\)}, where \(\tilde{n}_{even}\) is equal to
		\begin{align}
		\tilde{n}_{even}(x)=&\cos(\frac{\pi}{2}\frac{(x-L)}{w})\theta(w-\abs{x-L})+\cos(\frac{\pi}{2}\frac{(x+L)}{w})\theta(w-\abs{x+L})\\
		\tilde{n}_{even}(q)=&-\frac{2\pi w}{2L_{0}}\frac{\cos(kw)\cos(kL)}{k^2w^2-\qty(\pi/2)^{2}}
		\end{align}

\end{widetext}
	\begin{figure*}[tbp]
		\includegraphics[width=\linewidth]{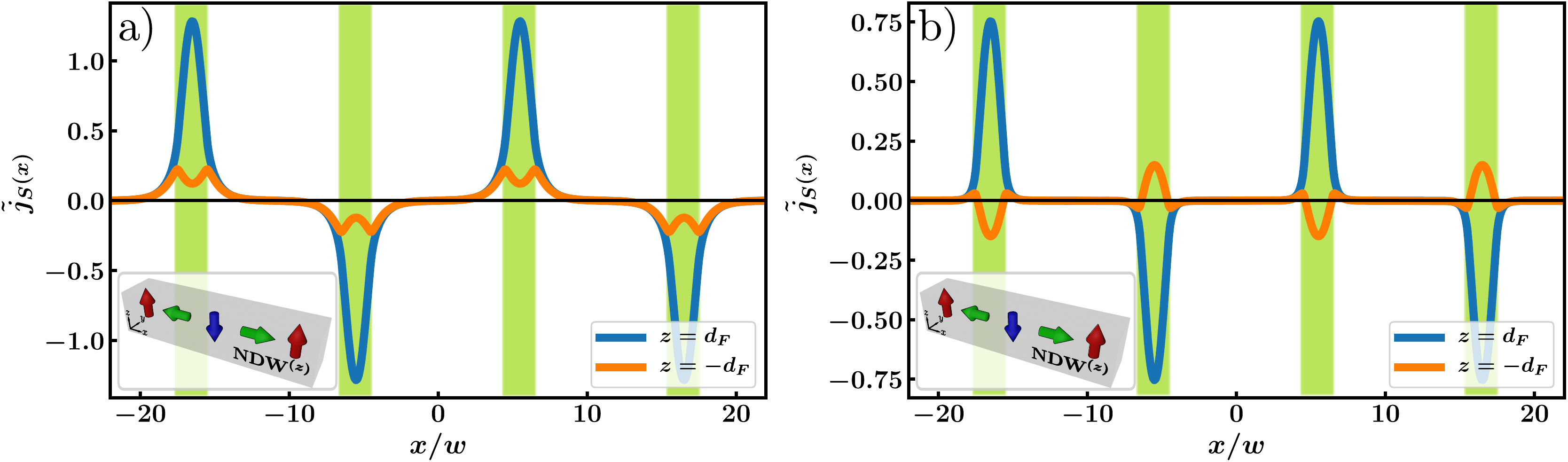}
		\caption{(Color online) a) Spatial dependence of the normalized Meissner current \mbox{\(\tilde{j}_{\text{S}}=2L_{0}j_{\text{S}}/(4\pi M_{0}cw^{2}\lambda_{\text{L}}^{-2})\)} for a periodic NDW$^{(z)}$ for \(L_{0}=11w\), \(L=(11/2)w\), \(\lambda_{\text{L}}=w\) and \(d_{\text{F}}=0.2w\). The structure of a single period \(2L_{0}\) of the NDW$^{(z)}$ is shown in the insets. The currents flow mainly above/below the DW regions, which are indicated by the green areas in the figure. The curents above and below the superconductor show a strong asymmetric behavior. b) Meissner current for the same configuration but with \(\lambda_{\text{L}}=(1/3)w\).  In this case the asymmetry can even lead to sign changes across the domain.   \label{Fig: NDW}}
	\end{figure*}
In order to obtain our final result for the magnetic stray-field and the Meissner current from Eq.(\ref{Eq: DW HSx}-\ref{Eq: DW jS}), we need to determine the constant \(C_{x}\). As mentioned in \mbox{Sec. \ref{Section1}}, the average over the in-plane component \(H_{x}^{(\text{F})}\) has to vanish, \textit{i.e}, \mbox{\(\expval{H_{x}^{(\text{F})}(x,z)}=0\)}. From this follows
	\begin{align}
		\expval{H_{x}^{(\text{F})}(x,z)}=&\frac{1}{2L_{0}}\int_{-L_{0}}^{L_{0}}\dd{x}\frac{1}{2d_{\text{F}}}\int_{-d_{\text{F}}}^{d_{\text{F}}}\dd{z}H_{x}^{(\text{F})}(x,z)\\
		=&\frac{4\pi M_{0}\lambda_{\text{L}}^{-1}d_{\text{F}}}{1+\lambda_{\text{L}}^{-1}d_{\text{F}}}\qty[\mathrm{n}_{x}(k=0)-C_{x}]\overset{!}{=}0
	\end{align}
where we used \mbox{\(\int_{-L_{0}}^{L_{0}}\dd{x}H_{x}^{(\text{F})}(x,z)=2L_{0}H_{x}^{(\text{F})}(k=0,z)\)}. Hence, the constant \(C_{x}\) is given by
\begin{equation}
		C_{x}=\mathrm{n}_{x}(k=0,z)
\end{equation}
The quantity \mbox{\(\mathrm{n}_{x}(k=0,z)\)} can be either \mbox{\(n_{even}(k=0,z)\)} or \mbox{\(n_{odd}(k=0,z)\)}. In the latter case, \(\mathrm{n}_{x}(k=0,z)=0\) (see Eq.(\ref{Eq:noddk},\ref{Eq:foddk})) and therefore \(C_{x}=0\). The other case is only realized for certain \mbox{N\'{e}el--type} DW's and results in
\begin{equation}
		C_{x}=\frac{2L}{L_{0}}-1
\end{equation}
\textit{i.e.}, the constant \(C_{x}\) vanishes for \(L_{0}=2L\). Otherwise,  if \mbox{\(L_{0}\neq2L\)},  the domains with positive and negative magnetization differ in size, which leads to an uncompensated total magnetization \(\vb{M}_{un}\) and \mbox{\(C_{x}\neq0\)}.
We will now examine the various possible magnetic textures that exhibit a chirality as defined in Eq.(\ref{Eq:noddx}). Note that, the type of DW in a ferromagnetic sample is determined by the existing magnetic interaction and material specific parameters (temperature, thickness of the F film etc). Accordingly, the actual magnetic texture in the F corresponds to the configuration associated with the minimum of the thermodynamic potential. Nevertheless, we will find the spatial distribution of the Meissner currents for all possible configurations, bearing in mind that some of these textures might not be  energetically favorable, but could be achieved in artificial magnetic structures \cite{Marioni_Halbach_2018}.

\subsection{Out-of-plane \(\vb{n}\) (N\'{e}el and Bloch DW's)}	
For an out-of-plane magnetization, both  N\'{e}el and/or  Bloch DW's (see \mbox{Fig. \ref{Fig: Schema}a and b)} can exist within the F. The N\'{e}el DW (NDW$^{(z)}$) is described by the following configuration \(\vb{n}(x)\)%
	\begin{equation}
		\vb{n}(x)=(n_{odd},0,n_{even}), \qquad \text{NDW}^{(z)}
	\end{equation}
	
	\begin{widetext}
The superscript \((z)\) indicates the alignment of the vector \(\vb{n}(x)\) across the domains, which is oriented along the \(z\)-axis. The Meissner current at the interfaces \mbox{\(z=\pm d_{\text{F}}\)} is obtain by inserting the corresponding Fourier components in Eq.(\ref{Eq: DW jS}).
	\begin{equation}
		j_{\text{S}}(k,\pm d_{\text{F}})=cM_{0}\lambda_{\text{L}}^{-2}\sinh(\abs{k}d_{\text{F}})\frac{2\pi iw}{2L_{0}}\qty[\pm\frac{f(k)}{D_{2}(k)}+\frac{\text{sgn}(k)F(k)}{D_{1}(k)}]
	\end{equation}
where \(f(k)\) and \(F(k)\) are given in Eq.(\ref{Eq:foddk},\ref{Eq:fevenk}). One can easily see that the current is an odd function of \(k\). In the coordinate representation, we obtain the following result
\begin{equation}
		j_{\text{S}}(x,\pm d_{\text{F}})=-\frac{4\pi M_{0}cw\lambda_{\text{L}}^{-2}}{2L_{0}}\sum_{k=0}^{\infty}\sin(kx)\sinh(k d_{\text{F}})\qty[\pm\frac{f(k)}{D_{2}(k)}+\frac{F(k)}{D_{1}(k)}]
\end{equation}
\end{widetext}
	\begin{figure*}[tbp]
		\includegraphics[width=\linewidth]{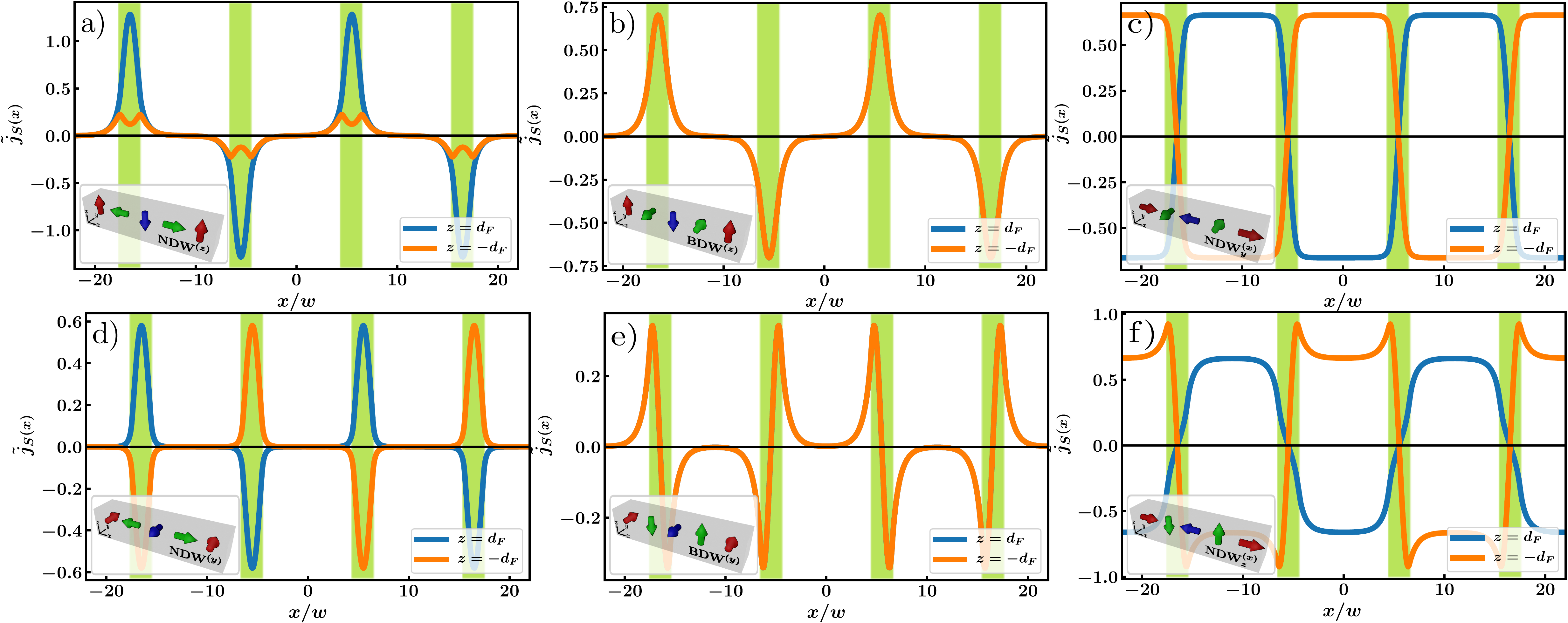}
		\caption{(Color online) Spatial dependence of the normalized Meissner current \mbox{\(\tilde{j}_{\text{S}}=2L_{0}j_{\text{S}}/(4\pi M_{0}cw^{2}\lambda_{\text{L}}^{-2})\)} for \(L_{0}=11w\), \(L=(11/2)w\), \(\lambda_{\text{L}}=w\) and \(d_{\text{F}}=0.2w\). The figures describe different periodic magnetic textures (see insets for the depiction of a single period \(2L_{0}\): \mbox{a) NDW$^{(z)}$}, \mbox{b) BDW$^{(z)}$}, \mbox{c)  NDW$^{(x)}_{y}$}, \mbox{d)  NDW$^{(y)}$}, \mbox{e) BDW$^{(y)}$} and \mbox{f) NDW$^{(x)}_{z}$}). The magnetization continuously changes via the DW's, which are indicated by the green areas. Depending on the underlying magnetization, screening currents preferably flow either above/below the DW, a), b), d) and e), or above/below the domain, c) and f). Their general distribution is unique for every DW structure. In the presence of Bloch--like magnetic textures, b) and e), the Meissner current in the upper and lower superconductors are identical. This is not the case for Neel-like structures, since they exhibit either antisymmetric, c) and d), or even asymmetrical behavior, a) and f). \label{Fig: Meissner}}
	\end{figure*}	
In Fig. \mbox{\ref{Fig: NDW}a} we plot the dependence of the normalized current \mbox{\(\tilde{j}_{\text{S}}=2L_{0}j_{\text{S}}/(4\pi M_{0}cw^{2}\lambda_{\text{L}}^{-2})\)} for \mbox{\(L_{0}=11w\)}, \mbox{\(L=(11/2)w\)}, \mbox{\(\lambda_{\text{L}}=w\)} and \mbox{\(d_{\text{F}}=0.2w\)}. This plot shows a strong asymmetry between the upper and lower superconductors with currents flowing above/below the DW regions. The direction of the supercurrent depends on the direction of rotation of the DW. Varying the value for the London penetration depth \mbox{\(\lambda_{\text{L}}=(1/3)w\)} reveals a sign change for the supercurrent within DW's in the lower supercurrent (see \mbox{Fig. \ref{Fig: NDW}b)}. One can see  that the Meissner currents at different DW's flow in opposite directions. This means that the currents flow along closed loops. Unlike the case of Abrikosov vortices, there is no phase change along these loops. This sign change is similar to the behavior described for the N\'{e}el Sk's. As in the case of the Sk, the asymmetry follows from the non-vanishing \(\div{\vb{M}}\) term inside the FM, which means that both the bulk and the surface magnetic charges are present. Note that in the zero-width domain wall limit, the \(\vb{n}_{x}\) component vanishes, \textit{i.e.}, \(\div{\vb{M}}=0\). As a result the asymmetry would disappear.
	
In the case of a Bloch DW (BDW$^{(z)}$) the vector \(\vb{n}(x)\) has the components
	\begin{equation}
		\vb{n}(x)=(0,n_{odd},n_{even}), \qquad \text{BDW}^{(z)}
	\end{equation}
where the magnetization in the domain is once again oriented along the \(z\)-direction. The Meissner current in Fourier representation is given by
	\begin{equation}
		j_{\text{S}}(k,\pm d_{\text{F}})=cM_{0}\lambda_{\text{L}}^{-2}\sinh(\abs{k}d_{\text{F}})\frac{2\pi iw}{2L_{0}}\frac{\text{sgn}(k)F(k)}{D_{1}(k)}
	\end{equation}
and in the coordinate representation by
	\begin{equation}
		j_{\text{S}}(x,\pm d_{\text{F}})=-\frac{4\pi M_{0}cw\lambda_{\text{L}}^{-2}}{2L_{0}}\sum_{k=0}^{\infty}\sin(kx)\sinh(k d_{\text{F}})\frac{F(k)}{D_{1}(k)}
	\end{equation}
One can directly deduce that the resulting Meissner currents are identical in the upper and lower superconductors, which is due to the missing \(x\)-component of the magnetization. This is once again, similar to the Sk case, as there was also no asymmetry present for Bloch Sk's. For both NDW$^{(z)}$ and BDW$^{(z)}$ follows that \(j_{\text{S}}(x,z)\) is an odd function of \(x\) so that the total current \mbox{\(J_{av}=\int\dd{x}j_{\text{S}}(x,\pm d_{\text{F}})\)} vanishes. In \mbox{Fig. \ref{Fig: Meissner}b}, we plot the dependence of the Meissner current for the considered case of a BDW$^{(z)}$. The parameter are the same as in \mbox{Fig. \ref{Fig: NDW}}. 

\subsection{In-plane \(\vb{n}\) (N\'{e}el and Bloch DW's)}

Let us first consider a \mbox{N\'{e}el--type} DW where the magnetization vector \(\vb{n}(x)\) at the domains is oriented along the \mbox{\(y\)-direction} (see \mbox{Fig. \ref{Fig: Schema}d}), then
	\begin{equation}
		\vb{n}(x)=(n_{odd},n_{even},0), \qquad \text{NDW}^{(y)}
	\end{equation}
	The Fourier component of the Meissner current is equal to
	\begin{equation}
		j_{\text{S}}(k,\pm d_{\text{F}})=\pm cM_{0}\lambda_{\text{L}}^{-2}\sinh(\abs{k}d_{\text{F}})\frac{2\pi iw}{2L_{0}}\frac{f(k)}{D_{2}(k)}
	\end{equation}
	and in the coordinate representation
	\begin{equation}
		j_{\text{S}}(x,\pm d_{\text{F}})=\mp\frac{4\pi M_{0}cw\lambda_{\text{L}}^{-2}}{2L_{0}}\sum_{k=0}^{\infty}\sin(kx)\sinh(k d_{\text{F}})\frac{f(k)}{D_{2}(k)}
	\end{equation}
The functions \(j_{\text{S}}(x,d_{\text{F}})\) and \(j_{\text{S}}(x,-d_{\text{F}})\) are shown in \mbox{Fig. \ref{Fig: Meissner}d}. Once again the currents differ in the two superconducting regions. The magnitude of the currents is the same, but the currents flow in opposite direction, resulting in an antisymmetric behavior. 
	
The Bloch type DW (see \mbox{Fig. \ref{Fig: Schema}e}) is described by
	\begin{equation}
		\vb{n}(x)=(0,n_{even},n_{odd}), \qquad \text{BDW}^{(y)}
	\end{equation}
	The occurring currents \(j_{\text{S}}(x,z)\) for BDW$^{(y)}$ are even function of \(x\) given by
	\begin{equation}
		j_{\text{S}}(x,\pm d_{\text{F}})=-\frac{4\pi M_{0}cw\lambda_{\text{L}}^{-2}}{2L_{0}}\sum_{k=0}^{\infty}\cos(kx)\sinh(k d_{\text{F}})\frac{f(k)}{D_{1}(k)}
	\end{equation}
    For both \mbox{\(z=\pm d_{\text{F}}\)} the currents are equal. In \mbox{Fig. \ref{Fig: Meissner}e}, we plot the \(x\)-dependence of the functions \(j_{\text{S}}(x,d_{\text{F}})\) and \(j_{\text{S}}(x,-d_{\text{F}})\). The Meissner currents in the upper and lower S near the BDW flow in the same direction. The total current \(j_{av}\) is zero. The results for \(j_{\text{S}}(x,z)\) for the BDW$^{(y)}$ are similiar to those obtained by Burmistrov and Chtchelkatchev\cite{Burmistrov2005}.

    \subsection{Other types of NDW}
    Other types of NDW's correspond to a magnetization profile \(\vb{n}(x)\) in which the alignment in the domain is along the \(x\)-direction (see \mbox{Fig. \ref{Fig: Schema}c and f}). The rotation of the vector \(\vb{n}(x)\) occurs either in the \((x,z)\)-plane or in the \((x,y)\)-plane. Thus, the vector \(\vb{n}(x)\) has the components 
    \begin{align}
    	\vb{n}(x)=&(n_{even},n_{odd},0), \qquad \text{NDW}_{y}^{(x)}\\
    	\vb{n}(x)=&(n_{even},0,n_{odd}), \qquad \text{NDW}_{z}^{(x)}
    \end{align}
    Remember, that for the case \mbox{\(n_{x}(x)=n_{even}(x)\)} the constant \(C_{x}\) has a finite value given by \mbox{\(C_{x}=2L/L_{0}-1\)}. It follows that 
    \begin{equation}
   		\mathrm{\bar{n}}_{x}(k)=n_{even}(x)-C_{x}\frac{\sin(kL_{0})}{kL_{0}}=\frac{2\pi w}{2L_{0}}\bar{F}(k)
    \end{equation}
    with 
    \begin{equation}
	 	\bar{F}(k)=\frac{\pi}{2kw}f(k)-\frac{2L}{L_{0}}\frac{\sin(kL_{0})}{\pi kw}
    \end{equation}
    
    \begin{widetext}
    With this expression, we obtain the Meissner current
    \begin{align}
    	j_{\text{S}}^{1}(x,\pm d_{\text{F}})=&\pm\frac{4\pi M_{0}cw\lambda_{\text{L}}^{-2}}{2L_{0}}\sum_{k=0}^{\infty}\cos(kx)\sinh(k d_{\text{F}})\frac{\bar{F}(k)}{D_{2}(k)}\\
    	j_{\text{S}}^{2}(x,\pm d_{\text{F}})=&\frac{4\pi M_{0}cw\lambda_{\text{L}}^{-2}}{2L_{0}}\sum_{k=0}^{\infty}\cos(kx)\sinh(k d_{\text{F}})\qty[\pm\frac{\bar{F}(k)}{D_{2}(k)}-\frac{f(k)}{D_{1}(k)}]
    \end{align}  
    
    \end{widetext}
    
	We continue to consider a compensated magnetization where \mbox{\(L_{0}=2L\)} so that \mbox{\(\bar{F}(k)=F(k)\)}.\\
	Once again, we find the typical asymmetry associated with \mbox{N\'{e}el--type} magnetic textures. The plots for the two magnetization NDW$_{y}^{(x)}$ and NDW$_{z}^{(x)}$ can be found in \mbox{Fig. \ref{Fig: Meissner}c} and \mbox{Fig. \ref{Fig: Meissner}f}, respectively.
	
	\section{Nucleation of superconductivity\label{Section4}}
	
In this section, we analyze qualitatively the nucleation of superconductivity in S films in an S/F/S structure when the temperature drops below the critical value in a bulk superconductor \(T_{cB}\). The value of the critical temperature \(T_{c}\) differs from its bulk value in the presence of a local depairing factor \(V(\vb{r}_{\perp})\). This problem was analyzed in the case of ferromagnetic superconductors with DW's both theoretically \cite{Melnikov_2003,Moshchalkov_2005,Moshchalkov_2006,Silaev_2014} and experimentally\cite{Bader_2003,Moshchalkov_2005,Norman_2006,Moshchalkov_2006}. Results of intensive studies of hybrid S/F structures are summarized in a review\cite{Aladyshkin_2009}. Theoretical studies were carried out under approximation of zero DW width. In particular the authors of Ref.\cite{Silaev_2014} have studied the so-called diode effect, that is an asymmetrical dependence of the critical current \(\vb{j}_{cr}\) in the S film with respect to the "magnetic current" \(\vb{j}_{M}=\curl{\vb{M}}\). Near the critical temperature at which superconductivity nucleates or disappears, the Ginzburg-Landau equation for the order parameter (OP) \(\psi\)  can be linearized. In the studies\cite{Yu_2003,Moshchalkov_2005,Moshchalkov_2006,Silaev_2014}, the OP \(\psi\) is presented in the form \(\psi=f\exp(i\chi)\), where in a one-dimensional case the phase of the OP is \(\chi=kx\). However, in a single-connected superconductor and in the absence of vortices one can choose a gauge with \(\chi=0\). On the other hand, if we consider our system in the form of a ring (annular geometry) with \(x=r\varphi\), then the phase is \(\chi=n\varphi\), where \(\varphi\) is the azimuthal angle, n is an integer number of fluxons and \(R\) is the radius of the ring. Even in this case the gauge is not established in a unique way: one can add  in principle an arbitrary constant \(\vb{A}_{ext}\) to the vector potential \(\vb{A}\) without changing the observable quantity, the magnetic induction \(\vb{B}=\curl{\vb{A}}\). From the physical point of view, adding a constant \(\vb{A}_{ext}\) means adding a uniform external or spontaneous (in the annular setup) current.
Observe that calculating the spatial dependence \(A(\vb*{r}_{\perp})\), we assumed, unlike previous theoretical studies, that \(\vb{A}_{ext}=0\). This means that the total current in the system vanishes \(\expval{j_{\text{S}}(\vb*{r}_{\perp})}=-(c/4\pi)\expval{A(\vb*{r}_{\perp})}=0\).

\begin{widetext}	
Thus, assuming \(A_{ext}=0\), we consider the Ginzburg-Landau equations for the order parameter f  (see, for example, Ref. \cite{Aladyshkin_2006}) near $T_c$.
	\begin{equation}
		-d_{\text{S}}\nabla_{\vb*{r}_{\perp}}^{2}f+V(\vb*{r}_{\perp})f-Ef=-Ef^{3}\label{Eq: GL}
	\end{equation}
where \mbox{\(V(\vb*{r}_{\perp})=d_{\text{S}}(2\pi \{A_{0}(\vb*{r}_{\perp})-A_{ext}\}/\Phi_{0})^2\)}. Here \(A_{0}(\vb*{r}_{\perp})\) is not a gauge invariant quantity, so any observable quantity like the magnetic induction \(\vb{B}=\curl{\vb{A}}\) will not change by adding a constant \(A_{ext}\). However, in doing so it is necessary to analyze the excitation mechanism of the persistent current and to estimate the condensation energy \(E_{\text{S}}=\Delta^{2}N(0)V\) in comparison to the magnetic energy \(F_{I}=I_{ext}^{2}\mathcal{L}/c^{2}\) of the current \(I_{ext}\), where \(N(0)\) is the density-of-states in the normal state, \(V=2\pi R(2_{\text{S}})L_{x}\) is the volume of the superconductors and \(\mathcal{L}\) is the inductance of the superconducting ring.  The ratio \(\gamma\) of these two energies depends on various parameters of the system such as the London penetration depth \(\lambda_{\text{L}}\) and the DW width \(w\) and may be both smaller or larger than 1. Thus, the excitation of the current \(I_{ext}\) may not be necessarily energetically favorable. In addition, for a single-connected superconducting system, the problem should be then solved under the condition of the absence of the total current in a finite S/F/S structure disconnected from external circuits. Our solution with $A_{ext}=0$ would immediately satisfy these constrains. A more detailed analysis of the problem for a finite current \(I_{ext}\) should be left for future studies.   

The vector potential \(A_{0}\) defines the stray-field in absence of superconductivity, which can be extracted from Eq.(\ref{Eq: Vectorpotential S}) by taking the limit \(\lambda_{\text{L}}\rightarrow\infty\). 
\begin{equation}
\vb{A}_{\text{0}}(\vb*{k},z)=\frac{4\pi M_{0}}{\abs{k}}\sinh(\abs{k}d_{\text{F}})\qty(\hat{\vb{e}}_{z}\times\frac{i\vb*{k}}{\abs{k}})\qty[\pm\frac{i\vb*{k}}{\abs{k}}\vb{n}_{\perp}(\vb*{k})-\mathrm{n}_{z}(\vb*{k})]e^{-\abs{k}\abs{z}}
\end{equation}
%

For simplicity, we assume that the thickness of the S films \(d_{\text{S}}\) is smaller than \(\xi_{\text{S}}\), so that the order parameter (OP) \(f\) depends only on the in-plane  coordinates. At larger \(d_{\text{S}}\), the factor \(V(\vb*{r}_{\perp},z)\) depends on the coordinate \(z\) and the effect of this depairing factor on the nucleation of superconductivity becomes weaker. Eq.(\ref{Eq: GL}) is called the time-independent Gross-Pitaevskii \cite{Gross_1961,pitaevskii_1961} equation or nonlinear Schr\"odinger equation.  The "energy" \(E\) is related to the coherence length \mbox{\(\xi_{\text{S}}(T)=\xi_{\text{S0}}/\sqrt{1-T/T_{cB}}\)}, \mbox{\(E=d_{\text{S}}/\xi_{\text{S}}^{2}\)}. This equation is also used to analyze the nucleation of superconductivity near the critical magnetic field \(H_{c2}\) (see Abrikosov's book \cite{abrikosov_fundamentals_1988} and \cite{abrikosov_magnetic_1957}) and also near DW in a S/F system \cite{Aladyshkin_2006}. Note also Ref.\cite{Moor_2014}, where this equation is applied for studying the appearance of an OP in a system with two competing OPs.

In the following, we will focus on DW structures where in Eq.(\ref{Eq: GL}) \mbox{\(\nabla_{\vb*{r}_{\perp}}\rightarrow(\partial_{x},0)\)}. In this case the real space expression for the vector potential at the interface \mbox{\(z=\pm d_{\text{F}}\)} is given by
	\begin{equation}
		\vb{A}_{0}(x,\pm d_{\text{F}})=-4\pi M_{0}\sum_{k}\frac{\sinh(\abs{k}d_{\text{F}})}{\abs{k}}e^{ikx}\qty[\pm \mathrm{n}_{x}(k)+\frac{ik}{\abs{k}}\mathrm{n}_{z}(k)]\vb{\hat{e}}_{y}
	\end{equation}
	\end{widetext}

\begin{figure*}[tbp]
			\includegraphics[width=\linewidth]{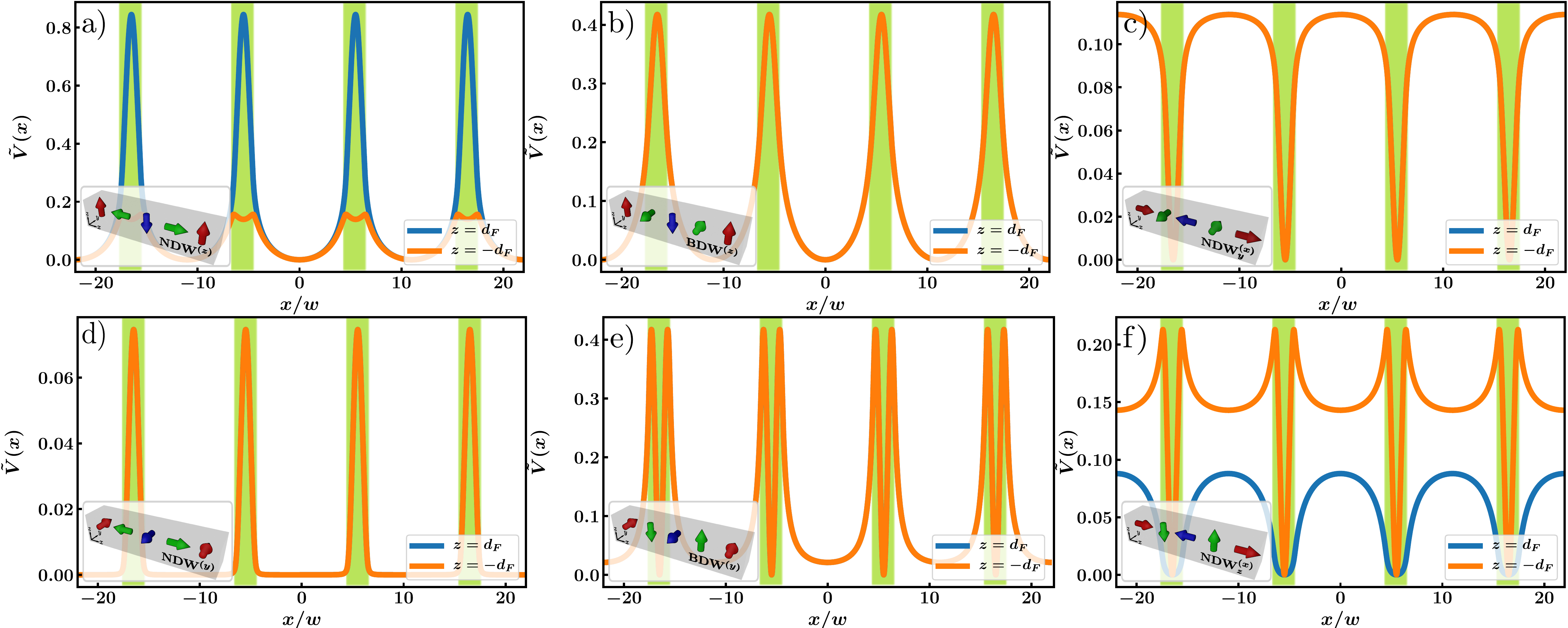}
			\caption{(Color online) Spatial dependence of the normalized depairing potential \mbox{\(\tilde{V}(x)=d_{\text{S}}V(x)(2L_{0}\Phi_{0}/8\pi^{2} M_{0}w^{2})^{2}\)} for \mbox{\(L_{0}=11w\)}, \mbox{\(L=(11/2)w\)} and \mbox{\(d_{\text{F}}=0.2w\)}. The figures describe different periodic magnetic textures (see insets for the depiction of a single period \(2L_{0}\): \mbox{a) NDW$^{(z)}$}, \mbox{b) BDW$^{(z)}$}, \mbox{c)  NDW$^{(x)}_{y}$}, \mbox{d)  NDW$^{(y)}$}, \mbox{e) BDW$^{(y)}$} and \mbox{f) NDW$^{(x)}_{z}$}), where the magnetization continuously changes across the green areas indicating the DW regions. The minima of the potential are either at the DW's, for  c), e), f), or in the center of the domain, for a), b), d), and correspond to the regions where superconductivity nucleates first. In contrast to the Meissner current (see \mbox{Fig. \ref{Fig: Meissner}}), there is no antisymmetric behavior for NDW$^{(x)}_{y}$ (c) and NDW$^{(y)}$ (d), so that the potential in the upper and lower superconductor is the same. In the case of NDW$^{(z)}$ (a)) and NDW$^{(x)}_{y}$ (f)), however, the potential remains asymmetric, which should lead to different critical temperatures in the superconductors.\label{Fig: Potential}}
	\end{figure*}

The associated depairing potential \mbox{\(V(x)\propto A_{0}^{2}(x)\)} is shown in \mbox{Fig. \ref{Fig: Potential}} for the different magnetization configurations. The potential \(V(x)\) has minima located either at the DW's \mbox{(\(x=(2n+1)L,\quad n=0,\pm1,..\))} or in the center of the domains \mbox{(\(x=2nL,\quad n=0,\pm1,..\))}. The critical temperature \(T_{c}\) is determined by the condition \mbox{\(E_{min}=d_{\text{S}}/\xi_{\text{S}}^{2}(T_{c})\)}, where \(E_{min}\) is the minimal "energy" at which Eq.(\ref{Eq: GL}) has a non-trivial solution. We assume that the domain size \(2L\) is much larger than the width of the DW \(2w\). In the following we consider two possible cases.
	
	\subsubsection{\(V(x)\) has a minimum at the DW}
	Consider first the case when the potential \(V(x)\) has a sharp minimum at the DW, for example, at \mbox{\(\tilde{x}=0\)} where \mbox{\(\tilde{x}=x-L\)}. Since we are interested in a qualitative picture, we approximate the dependence \(V(x)\) near the DW with a rectangular potential well: \mbox{\(V(x)=V_{0}-V_{0}\theta(w-\abs{\tilde{x}})\)}. Then, neglecting the r.h.s. in Eq.(\ref{Eq: GL}) and using the matching conditions at \mbox{\(x=\pm w\)} (\(f(x)\) and \(\partial_{x}f(x)\) are continuous) we can write a solution in the form
	\begin{equation}
		f(x)=C_{in}\cos(K_{\text{S}}\tilde{x}),\qquad \abs{\tilde{x}}<w
	\end{equation} 
	\begin{equation}
		f(x)=C_{out}\begin{cases}
		\exp(-K_{out}(\tilde{x}-w)), \qquad &\tilde{x}> w\\
		\exp(-K_{out}(\tilde{x}+w)), \qquad &\tilde{x}< -w
		\end{cases}
	\end{equation}
	where \mbox{\(K_{\text{S}}=\xi_{\text{S}}^{-2}(T_{c})\)}, and \mbox{\(K_{out}=(V_{0}/d_{\text{S}})-\xi_{\text{S}}^{-2}(T_{c})\)}. The integration constants \(C_{in}\) and \(C_{out}\) are related to each other. In the limiting cases of small and large \mbox{\(\lambda\equiv\sqrt{V_{0}w^{2}/d_{\text{S}}}\)}, we have for \(T_{c}\) and \(C_{in}\), \(C_{out}\)
	
\noindent	a) \mbox{\(\lambda\ll1\); \(C_{out}\approx C_{in}\) and \(T_{c}\approx T_{cB}(1-V_{0}\xi_{\text{S0}}^{2}/d_{\text{S}})\)}; \\
\noindent	b) \mbox{\(\lambda\gg1\); \(C_{out}\approx C_{in}/\lambda^{2}\) and \(T_{c}\approx T_{cB}(1-(\pi\xi_{\text{S0}}/w)^{2})\)}.\\

Thus, if the depairing potential \(V(x)\) has a dip at the DW's, superconductivity is nucleated at the DW's. This happens in the following DW configurations: BDW$^{(y)}$, NDW$_{y}^{(x)}$, NDW$_{z}^{(x)}$ (the potentials are shown in \mbox{Fig. \ref{Fig: Potential}}). The opposite case is realized for the magnetization profiles:  NDW$^{(z)}$, BDW$^{(z)}$, NDW$^{(y)}$ (see also \mbox{Fig. \ref{Fig: Potential}} for the respecting potentials) and is considered in the next section.

For simplicity, we neglect the width \(w\) in comparison with \(L\). The solution outside the DW has the form
	\begin{equation}
		f(x)=C_{out}\cos(K_{\text{S}x}), \qquad \abs{x}<L
	\end{equation}
The critical temperature \(T_{c}\) is found from the matching condition \mbox{\([\partial_{x}f]_{x=\pm L}=V(x)f(x)|_{x=\pm L}\)}. The constant \mbox{\(C=C_{out}\)} is not zero provided that the condition
	\begin{equation}
		\theta_{c}\tan(\theta_{c})=\lambda
	\end{equation}
is fulfilled, where \mbox{\(\theta_{c}=L/\xi_{\text{S}}(T_{c})\)}. Eq.(\ref{Eq: GL}) yields for the critical temperature \(T_{c}\):\\
\noindent a) \mbox{\(T_{c}/T_{cB}=1-\lambda(\xi_{\text{S}0}/L)^{2}\) for \(\lambda\ll 1\)}\\
\noindent b) \mbox{\(T_{c}/T_{cB}=1-\lambda(\pi\xi_{\text{S}0}/2L)^{2}\) for \(\lambda\gg 1\)}\\
	The constant \(C\) is found analogously to the case considered in Ref.\cite{Moor_2014}
	\begin{equation}
		C^{2}=\frac{T-T_{c}}{T_{cB}-T_{c}}r(\theta_{L})
	\end{equation}
	with
	\begin{equation}
		r(\theta_{L})=\frac{\expval{\cos^{2}(K_{\text{S}}x)}}{\expval{\cos^{4}(K_{\text{S}}x)}}=\frac{2\theta_{L}+\sin(2\theta_{L})}{(3/2)\theta_{L}+\sin(2\theta_{L})+(1/8)\sin(4\theta_{L})}
	\end{equation}
	where \mbox{\(\expval{...}=\int_{0}^{L}\dd{x}(...)\)} and \mbox{\(\theta_{L}=K_{\text{S}}L\)}. For \mbox{\(\theta_{L}\gg1\)} the coefficient \(r\) is equal to: \mbox{\(r=4/3\)}.
	\subsubsection{\(V(x)\) has a mimimum at the center of the domain}
	We assume that the potential \(V(x)\) in Eq.(\ref{Eq: GL}) has the form
	\begin{equation}
		V(x)=V_{0}\sum_{n}\delta(x-2nL)
	\end{equation}
	First, we linearize Eq.(\ref{Eq: GL}) and find the minimum "energy" \(E\) of the Schr\"odinger like equation in the interval \mbox{\(x\in{-L,L}\)}
	\begin{equation}
		-\partial_{\tilde{x}\tilde{x}}^{2}f+V_{0}\delta(\tilde{x})f(0)=\tilde{E}f
	\end{equation}
	where \(\tilde{x}=x/\xi_{\text{S}0}\), \(\tilde{E}=E(\xi_{\text{S}0}/d_{\text{S}})=1-T/T_{cB}\). A periodic solution (\(f(\tilde{x}=f(\tilde{x}+2L/\xi_{\text{S}0}))\)) can be represented in the form
	\begin{equation}
		f(\tilde{x})=\begin{cases}
		a\cos(q\tilde{x})+b\sin(q\tilde{x}),\qquad &0<\tilde{x}<L/\xi_{\text{S}0}\\
		\bar{a}\cos(q\tilde{x})+\bar{b}\sin(q\tilde{x}),\qquad &-L/\xi_{\text{S}0}<\tilde{x}<0 \label{Eq: f center domain}
		\end{cases}
	\end{equation}
	where \mbox{\(q^{2}=E\)}. The function \(f(\tilde{x})\) has to fulfill the matching conditions
	\begin{align}
		[f]= 0&& [\partial_{\tilde{x}}f]=V(\tilde{x})f(0)\notag\\
		f(L)=f(-L) && \partial_{\tilde{x}}f(\tilde{x})|x=L= \partial_{\tilde{x}}f(\tilde{x})|x=-L
	\end{align}
	The solution (\ref{Eq: f center domain}) exists if the condition
	\begin{equation}
		\theta\tan(\theta)=v \equiv VL/2 \label{Cond: Center domain}
	\end{equation}
	is satisfied where \mbox{\(\theta=qL\)}. The coefficients \(a\) and b are coupled by the relations: \mbox{\(a=\bar{a}\)}, \mbox{\(b=-\bar{b}=Va/2q\)}. From Eq.(\ref{Cond: Center domain}) we find the critical temperature
	\begin{equation}
		T_{c}/T_{cB}=\begin{cases}
		1-(v\xi_{\text{S}0}/L)^{2},\qquad &v\ll 1\\
		1-(\pi\xi_{\text{S}0}/2L)^{2},\qquad &v\gg 1
		\end{cases}
	\end{equation}
	If \mbox{\(L/\xi_{\text{S}0}>2/\pi\)}, superconductivity is suppressed completely.
	
	\section{Conclusion\label{Section5}}
	To conclude, in this manuscript we calculated the magnetic stray-field \(\vb{H}_{str}\) and Meissner current \(\vb{j}_{\text{S}}\) in a superconductors S created by various non-homogeneous magnetic texture in a F film incorporated in an S/F/S system. The total current in the system is assumed to be zero. Two types of topological structures were considered: isolated chiral magnetic skyrmions and periodic flat domain walls of Bloch (BDW) or \mbox{N\'{e}el--type} (NDW). Considering a two-dimensional two-component magnetization \(\vb{M}(\vb*{r}_{\perp})\), we investigated six different magnetic DW textures as well as magnetic Sk of Bloch and \mbox{N\'{e}el--type}. Each of these different magnetic textures possesses a particular spatial dependence of the stray-field \(\vb{H}_{\text{str}}(\vb*{r}_{\perp},z)\) and the induced Meissner current \(\vb{j}_{\text{S}}(\vb*{r}_{\perp},z)\).
	The most apparent difference appears between the Bloch-- and the \mbox{N\'{e}el--type} magnetic structures. While the Neel-type structure yields a strong asymmetry \(j_{\text{S}}(\vb*{r}_{\perp},z)\neq j_{\text{S}}(\vb*{r}_{\perp},-z)\), the \mbox{Bloch--type} remains always symmetric w.r.t \(z\)-component.  For certain parameter, this asymmetry can be strong enough to cause a sign change of the Meissner current for \(\vb*{r}_{\perp}\) within the DW region or within the Sk radius \(\vb*{r}_{\perp}\).
	Note that a similar sign change can be obtained in S/F or S/F/S systems that feature a proximity effect\cite{bergeret_spin_2004,volkov_spin_2019,mironov_electromagnetic_2018}. 
	
	The Meissner current \(j_{\text{S}}\) is connected to the vector potential \(A\) via \mbox{\(j_{\text{S}}=c\lambda_{\text{L}}^{-2}A/4\pi\)} which enters the Ginzburg-Landau equation and acts as a depairing factor \mbox{\(V(\vb*{r}_{\perp})=2\pi A_{0}^{2}(\vb*{r}_{\perp})\)} where \(A_{0}\) is the vector potential in absence of superconductivity. This factor determines the critical temperature of the superconducting transition in bulk superconductors\cite{abrikosov_fundamentals_1988} and in S/F heterostructures\cite{Yu_2003,Burmistrov2005,Aladyshkin_2006} and the superconductivity emerges first at places where \(A_{0}\) has a minimum. As can be seen in \mbox{Fig. \ref{Fig: Potential}} the locations of the minima or maxima of \mbox{\(V\propto A_{0}^{2}\)} depends on the type of DW's. For magnetic skyrmions the depairing potential \(V(\vb*{r}_{\perp})\) has its minimum in the center of the Sk. However, it can exhibit an additional local minimum for finite \(\vb*{r}_{\perp}\) within the radius of a N\'{e}el Sk which is not present for \mbox{Bloch--type} skyrmions. 
	Thus, by measuring the location of the superconducting nucleation like it was done previously \cite{Iavarone2014}, one can determine the type of the DW or distinguish between Bloch-- and \mbox{N\'{e}el--type} skyrmions.
    	
   	\section{Acknowledgement} The authors acknowledge support from the Deutsche Forschungsgemeinschaft Priority Program SPP2137,
   	Skyrmionics, under Grant No. ER 463/10.

	\bibliography{literature} 
	
	\onecolumngrid
	\appendix
	\setcounter{equation}{0}
	\setcounter{figure}{0}
	\setcounter{table}{0}
	\makeatletter
	\renewcommand{\theequation}{A\arabic{equation}}
	\renewcommand{\thefigure}{A\arabic{figure}}
	\renewcommand{\bibnumfmt}[1]{[A#1]}


\end{document}